\documentstyle[10pt,epsf,epsfig,hangcaption,xspace,amssymb,amsfonts,amsmath,amsthm,cite,
dp_delphititle,lineno]{dp_delphi}
\makeindex
\def\DpPaperGroup{EP}
\def\DpPaperRef{2003-089}
\def\DpDate{6 October 2003}
\def\DpAuthors{DELPHI Collaboration}
\def\DpSubmit{(Accepted by Euro. Phys. J. C)}
\def\DpTitle{{  Search for \boldmath $\mbox{B}^0_s$-$\overline{\mbox{B}^0_s}$ 
oscillations \\
  in DELPHI using high-$p_t$ leptons}}
\def\DpComment{   }
\def\DpEMail{  }
\include{newcom}

\newcommand{\epsb}{\epsilon_b^{tag}}

\newcommand{\dmq}{\Delta m_q}
\newcommand{\dmd}{\Delta m_d}
\newcommand{\dms}{\Delta m_s}

\newcommand{\Bsb}{\overline{\mbox{B}^0_s}}
\newcommand{\Bs}{\mbox{B}^0_s}

\newcommand{\Bm}{\mbox{B}^{-}}

\newcommand{\Bq}{\mbox{B}^{0}_{q}}
\newcommand{\Bdb}{\overline{\mbox{B}^{0}_{d}}}

\newcommand{\Ds}{\mbox{D}_s}

\newcommand{\GeV}{\rm{GeV}}

\newcommand{\mumu}{\ifmmode {\mu^+\mu^-} \else ${\mu^+\mu^-} $ \fi}

\newcommand{\ba}{\begin{array}}
\newcommand{\ea}{\end{array}}
\newcommand{\bc}{\begin{center}}
\newcommand{\ec}{\end{center}}
\newcommand{\be}{\begin{equation}}
\newcommand{\enq}{\end{equation}}
\newcommand{\beq}{\begin{eqnarray}}
\newcommand{\eeq}{\end{eqnarray}}
\newcommand{\bes}{\begin{eqnarray*}}
\newcommand{\ees}{\end{eqnarray*}}
\newcommand{\Kz}{\ifmmode {\rm K^0_s} \else ${\rm K^0_s} $ \fi}
\newcommand{\Zz}{\ifmmode {\rm Z^0} \else ${\rm Z^0 } $ \fi}
\newcommand{\qqbar}{\ifmmode {\rm q\bar{q}} \else ${\rm q\bar{q}} $ \fi}
\newcommand{\ccbar}{\ifmmode {\rm c\overline{c}} \else ${\rm c\overline{c}} $ \fi}
\newcommand{\bbbar}{\ifmmode {\rm b\overline{b}} \else ${\rm b\overline{b}} $ \fi}
\newcommand{\xxbar}{\ifmmode {\rm x\bar{x}} \else ${\rm x\bar{x}} $ \fi}
\newcommand{\rphi}{\ifmmode {\rm R\phi} \else ${\rm R\phi} $ \fi}
\begin{document}
\makeatletter
\newcount\@tempcntc
\def\@citex[#1]#2{\if@filesw\immediate\write\@auxout{\string\citation{#2}}\fi
  \@tempcnta\z@\@tempcntb\m@ne\def\@citea{}\@cite{\@for\@citeb:=#2\do
    {\@ifundefined
       {b@\@citeb}{\@citeo\@tempcntb\m@ne\@citea\def\@citea{,}{\bf ?}\@warning
       {Citation `\@citeb' on page \thepage \space undefined}}%
    {\setbox\z@\hbox{\global\@tempcntc0\csname b@\@citeb\endcsname\relax}%
     \ifnum\@tempcntc=\z@ \@citeo\@tempcntb\m@ne
       \@citea\def\@citea{,}\hbox{\csname b@\@citeb\endcsname}%
     \else
      \advance\@tempcntb\@ne
      \ifnum\@tempcntb=\@tempcntc
      \else\advance\@tempcntb\m@ne\@citeo
      \@tempcnta\@tempcntc\@tempcntb\@tempcntc\fi\fi}}\@citeo}{#1}}
\def\@citeo{\ifnum\@tempcnta>\@tempcntb\else\@citea\def\@citea{,}%
  \ifnum\@tempcnta=\@tempcntb\the\@tempcnta\else
   {\advance\@tempcnta\@ne\ifnum\@tempcnta=\@tempcntb \else \def\@citea{--}\fi
    \advance\@tempcnta\m@ne\the\@tempcnta\@citea\the\@tempcntb}\fi\fi}
 
\makeatother
\begin{titlepage}
\pagenumbering{roman}
\CERNpreprint{\DpPaperGroup}{\DpPaperRef} 
\date{{\small\DpDate}} 
\title{\DpTitle} 
\address{\DpAuthors} 
\begin{shortabs} 
\noindent
%
\noindent
Oscillations in the $\mbox{B}^0_s$-$\overline{\mbox{B}^0_s}$ system
were studied in events selected from about 4.3 million hadronic Z$^0$ decays 
registered by DELPHI between 1992 and 2000.\\
This paper presents updates of two published analyses (\cite{ref:dslold,ref:lincl}).
The first analysis, which utilizes leptons emitted with
large momentum transverse to a jet, was improved by means
of a better algorithm for the vertex reconstuction
and a new algorithm for flavour-tagging at production time. 
The second analysis, which utilizes $\Ds$-lepton events,
was improved by optimizing the treatment of proper time resolution.\\
No signal of $\mbox{B}^0_{\mathrm s}$ oscillations was observed and limits on the 
mass difference between the physical $\mbox{B}^0_{\mathrm s}$ states were obtained
to be:
\begin{eqnarray*}
   \begin{array}{ll}
\dms > 8.0 ~ \mbox{ps}^{-1}~\mbox{at the 95\% C.L.} & \\
\mbox{\rm {with a sensitivity of}}~~ \dms = 9.1 ~ \mbox{ps}^{-1} 
   \end{array}
\end{eqnarray*}
in the high $p_t$ lepton analysis and
\begin{eqnarray*}
   \begin{array}{ll}
\dms > 4.9 ~ \mbox{ps}^{-1}~\mbox{at the 95\% C.L.} & \\
\mbox{\rm {with a sensitivity of}}~~ \dms = 8.6 ~ \mbox{ps}^{-1} 
   \end{array}
\end{eqnarray*}
in the $\Ds$-lepton analysis.\\
Previously published results on these analyses are superseed.

The combination of these results with those obtained 
in other independent analyses previously performed in DELPHI 
($\Ds$-hadron, exclusive $\Bs$, inclusive vertex) gives:
\begin{eqnarray*}
  \begin{array}{ll}
      \dms > 8.5 ~ \mbox{ps}^{-1}~\mbox{at the 95\% C.L.} & \\
   \mbox{\rm {with a sensitivity of}}~~ \dms = 12.0 ~ \mbox{ps}^{-1}.
  \end{array}
\end{eqnarray*}

\end{shortabs}
\vfill
\begin{center}
\DpSubmit \ \\ 
\DpComment \ \\
\DpEMail \ \\
\end{center}
\vfill
\clearpage
\headsep 10.0pt
\addtolength{\textheight}{10mm}
\addtolength{\footskip}{-5mm}
\begingroup
%
\newcommand{\DpName}[2]{\hbox{#1$^{\ref{#2}}$},\hfill}
\newcommand{\DpNameTwo}[3]{\hbox{#1$^{\ref{#2},\ref{#3}}$},\hfill}
\newcommand{\DpNameThree}[4]{\hbox{#1$^{\ref{#2},\ref{#3},\ref{#4}}$},\hfill}
\newskip\Bigfill \Bigfill = 0pt plus 1000fill
\newcommand{\DpNameLast}[2]{\hbox{#1$^{\ref{#2}}$}\hspace{\Bigfill}}
%
\footnotesize
\noindent
\DpName{J.Abdallah}{LPNHE}
\DpName{P.Abreu}{LIP}
\DpName{W.Adam}{VIENNA}
\DpName{P.Adzic}{DEMOKRITOS}
\DpName{T.Albrecht}{KARLSRUHE}
\DpName{T.Alderweireld}{AIM}
\DpName{R.Alemany-Fernandez}{CERN}
\DpName{T.Allmendinger}{KARLSRUHE}
\DpName{P.P.Allport}{LIVERPOOL}
\DpName{U.Amaldi}{MILANO2}
% ...
\normalsize
\endgroup
\titlefoot{Department of Physics and Astronomy, Iowa State
     University, Ames IA 50011-3160, USA
    \label{AMES}}
\titlefoot{Physics Department, Universiteit Antwerpen,
     Universiteitsplein 1, B-2610 Antwerpen, Belgium \\
     \indent~~and IIHE, ULB-VUB,
     Pleinlaan 2, B-1050 Brussels, Belgium \\
     \indent~~and Facult\'e des Sciences,
     Univ. de l'Etat Mons, Av. Maistriau 19, B-7000 Mons, Belgium
    \label{AIM}}
\titlefoot{Physics Laboratory, University of Athens, Solonos Str.
     104, GR-10680 Athens, Greece
    \label{ATHENS}}
\titlefoot{Department of Physics, University of Bergen,
     All\'egaten 55, NO-5007 Bergen, Norway
    \label{BERGEN}}
\titlefoot{Dipartimento di Fisica, Universit\`a di Bologna and INFN,
     Via Irnerio 46, IT-40126 Bologna, Italy
    \label{BOLOGNA}}
\titlefoot{Centro Brasileiro de Pesquisas F\'{\i}sicas, rua Xavier Sigaud 150,
     BR-22290 Rio de Janeiro, Brazil \\
     \indent~~and Depto. de F\'{\i}sica, Pont. Univ. Cat\'olica,
     C.P. 38071 BR-22453 Rio de Janeiro, Brazil \\
     \indent~~and Inst. de F\'{\i}sica, Univ. Estadual do Rio de Janeiro,
     rua S\~{a}o Francisco Xavier 524, Rio de Janeiro, Brazil
    \label{BRASIL}}
\titlefoot{Coll\`ege de France, Lab. de Physique Corpusculaire, IN2P3-CNRS,
     FR-75231 Paris Cedex 05, France
    \label{CDF}}
\titlefoot{CERN, CH-1211 Geneva 23, Switzerland
    \label{CERN}}
\titlefoot{Institut de Recherches Subatomiques, IN2P3 - CNRS/ULP - BP20,
     FR-67037 Strasbourg Cedex, France
    \label{CRN}}
\titlefoot{Now at DESY-Zeuthen, Platanenallee 6, D-15735 Zeuthen, Germany
    \label{DESY}}
\titlefoot{Institute of Nuclear Physics, N.C.S.R. Demokritos,
     P.O. Box 60228, GR-15310 Athens, Greece
    \label{DEMOKRITOS}}
\titlefoot{FZU, Inst. of Phys. of the C.A.S. High Energy Physics Division,
     Na Slovance 2, CZ-180 40, Praha 8, Czech Republic
    \label{FZU}}
\titlefoot{Dipartimento di Fisica, Universit\`a di Genova and INFN,
     Via Dodecaneso 33, IT-16146 Genova, Italy
    \label{GENOVA}}
\titlefoot{Institut des Sciences Nucl\'eaires, IN2P3-CNRS, Universit\'e
     de Grenoble 1, FR-38026 Grenoble Cedex, France
    \label{GRENOBLE}}
\titlefoot{Helsinki Institute of Physics, P.O. Box 64,
     FIN-00014 University of Helsinki, Finland
    \label{HELSINKI}}
\titlefoot{Joint Institute for Nuclear Research, Dubna, Head Post
     Office, P.O. Box 79, RU-101 000 Moscow, Russian Federation
    \label{JINR}}
\titlefoot{Institut f\"ur Experimentelle Kernphysik,
     Universit\"at Karlsruhe, Postfach 6980, DE-76128 Karlsruhe,
     Germany
    \label{KARLSRUHE}}
\titlefoot{Institute of Nuclear Physics,Ul. Kawiory 26a,
     PL-30055 Krakow, Poland
    \label{KRAKOW1}}
\titlefoot{Faculty of Physics and Nuclear Techniques, University of Mining
     and Metallurgy, PL-30055 Krakow, Poland
    \label{KRAKOW2}}
\titlefoot{Universit\'e de Paris-Sud, Lab. de l'Acc\'el\'erateur
     Lin\'eaire, IN2P3-CNRS, B\^{a}t. 200, FR-91405 Orsay Cedex, France
    \label{LAL}}
\titlefoot{School of Physics and Chemistry, University of Lancaster,
     Lancaster LA1 4YB, UK
    \label{LANCASTER}}
\titlefoot{LIP, IST, FCUL - Av. Elias Garcia, 14-$1^{o}$,
     PT-1000 Lisboa Codex, Portugal
    \label{LIP}}
\titlefoot{Department of Physics, University of Liverpool, P.O.
     Box 147, Liverpool L69 3BX, UK
    \label{LIVERPOOL}}
\titlefoot{Dept. of Physics and Astronomy, Kelvin Building,
     University of Glasgow, Glasgow G12 8QQ
    \label{GLASGOW}}
\titlefoot{LPNHE, IN2P3-CNRS, Univ.~Paris VI et VII, Tour 33 (RdC),
     4 place Jussieu, FR-75252 Paris Cedex 05, France
    \label{LPNHE}}
\titlefoot{Department of Physics, University of Lund,
     S\"olvegatan 14, SE-223 63 Lund, Sweden
    \label{LUND}}
\titlefoot{Universit\'e Claude Bernard de Lyon, IPNL, IN2P3-CNRS,
     FR-69622 Villeurbanne Cedex, France
    \label{LYON}}
\titlefoot{Dipartimento di Fisica, Universit\`a di Milano and INFN-MILANO,
     Via Celoria 16, IT-20133 Milan, Italy
    \label{MILANO}}
\titlefoot{Dipartimento di Fisica, Univ. di Milano-Bicocca and
     INFN-MILANO, Piazza della Scienza 2, IT-20126 Milan, Italy
    \label{MILANO2}}
\titlefoot{IPNP of MFF, Charles Univ., Areal MFF,
     V Holesovickach 2, CZ-180 00, Praha 8, Czech Republic
    \label{NC}}
\titlefoot{NIKHEF, Postbus 41882, NL-1009 DB
     Amsterdam, The Netherlands
    \label{NIKHEF}}
\titlefoot{National Technical University, Physics Department,
     Zografou Campus, GR-15773 Athens, Greece
    \label{NTU-ATHENS}}
\titlefoot{Physics Department, University of Oslo, Blindern,
     NO-0316 Oslo, Norway
    \label{OSLO}}
\titlefoot{Dpto. Fisica, Univ. Oviedo, Avda. Calvo Sotelo
     s/n, ES-33007 Oviedo, Spain
    \label{OVIEDO}}
\titlefoot{Department of Physics, University of Oxford,
     Keble Road, Oxford OX1 3RH, UK
    \label{OXFORD}}
\titlefoot{Dipartimento di Fisica, Universit\`a di Padova and
     INFN, Via Marzolo 8, IT-35131 Padua, Italy
    \label{PADOVA}}
\titlefoot{Rutherford Appleton Laboratory, Chilton, Didcot
     OX11 OQX, UK
    \label{RAL}}
\titlefoot{Dipartimento di Fisica, Universit\`a di Roma II and
     INFN, Tor Vergata, IT-00173 Rome, Italy
    \label{ROMA2}}
\titlefoot{Dipartimento di Fisica, Universit\`a di Roma III and
     INFN, Via della Vasca Navale 84, IT-00146 Rome, Italy
    \label{ROMA3}}
\titlefoot{DAPNIA/Service de Physique des Particules,
     CEA-Saclay, FR-91191 Gif-sur-Yvette Cedex, France
    \label{SACLAY}}
\titlefoot{Instituto de Fisica de Cantabria (CSIC-UC), Avda.
     los Castros s/n, ES-39006 Santander, Spain
    \label{SANTANDER}}
\titlefoot{Inst. for High Energy Physics, Serpukov
     P.O. Box 35, Protvino, (Moscow Region), Russian Federation
    \label{SERPUKHOV}}
\titlefoot{J. Stefan Institute, Jamova 39, SI-1000 Ljubljana, Slovenia
     and Laboratory for Astroparticle Physics,\\
     \indent~~Nova Gorica Polytechnic, Kostanjeviska 16a, SI-5000 Nova Gorica, Slovenia, \\
     \indent~~and Department of Physics, University of Ljubljana,
     SI-1000 Ljubljana, Slovenia
    \label{SLOVENIJA}}
\titlefoot{Fysikum, Stockholm University,
     Box 6730, SE-113 85 Stockholm, Sweden
    \label{STOCKHOLM}}
\titlefoot{Dipartimento di Fisica Sperimentale, Universit\`a di
     Torino and INFN, Via P. Giuria 1, IT-10125 Turin, Italy
    \label{TORINO}}
\titlefoot{INFN,Sezione di Torino, and Dipartimento di Fisica Teorica,
     Universit\`a di Torino, Via P. Giuria 1,\\
     \indent~~IT-10125 Turin, Italy
    \label{TORINOTH}}
\titlefoot{Dipartimento di Fisica, Universit\`a di Trieste and
     INFN, Via A. Valerio 2, IT-34127 Trieste, Italy \\
     \indent~~and Istituto di Fisica, Universit\`a di Udine,
     IT-33100 Udine, Italy
    \label{TU}}
\titlefoot{Univ. Federal do Rio de Janeiro, C.P. 68528
     Cidade Univ., Ilha do Fund\~ao
     BR-21945-970 Rio de Janeiro, Brazil
    \label{UFRJ}}
\titlefoot{Department of Radiation Sciences, University of
     Uppsala, P.O. Box 535, SE-751 21 Uppsala, Sweden
    \label{UPPSALA}}
\titlefoot{IFIC, Valencia-CSIC, and D.F.A.M.N., U. de Valencia,
     Avda. Dr. Moliner 50, ES-46100 Burjassot (Valencia), Spain
    \label{VALENCIA}}
\titlefoot{Institut f\"ur Hochenergiephysik, \"Osterr. Akad.
     d. Wissensch., Nikolsdorfergasse 18, AT-1050 Vienna, Austria
    \label{VIENNA}}
\titlefoot{Inst. Nuclear Studies and University of Warsaw, Ul.
     Hoza 69, PL-00681 Warsaw, Poland
    \label{WARSZAWA}}
\titlefoot{Fachbereich Physik, University of Wuppertal, Postfach
     100 127, DE-42097 Wuppertal, Germany
    \label{WUPPERTAL}}
\addtolength{\textheight}{-10mm}
\addtolength{\footskip}{5mm}
\clearpage
\headsep 30.0pt
\end{titlepage}
%
\pagenumbering{arabic} 
\setcounter{footnote}{0} %
\large
\section{Introduction}
\label{sec:1}

One of the main interests in B physics is the precise determination of the values of the 
$\rho$ and $\eta$ parameters, the two least known parameters
of the ${CKM}$ matrix~\cite{ref:PDG} in the Wolfenstein parameterisation.
The unitarity of this matrix can be visualized as a triangle in the $\rho-\eta$ plane.
Several quantities which depend on $\rho$ and $\eta$ can be measured and, if the Standard Model is
correct, they must yield, within measurement
errors and theoretical uncertainties, 
compatible values for the these two parameters, inside measurement
errors and theoretical uncertainties. 
One of these quantities is the mass difference ($\dmq$)
between CP eigenstates in the neutral B-meson systems ${\mathrm B}^0_{d(s)}$.

In the Standard Model, $\Bq-\overline{\Bq}$ ($q = d,s$) mixing is a
direct consequence of second-order weak interactions. Starting with a
 $\Bq$ meson produced at time $t$=0, the probability, ${\cal P}$, to observe a
 $\Bq$ decaying at the proper time $t$ can be written, neglecting effects from CP violation:
\bc
${\cal P}(\Bq\rightarrow \Bq)~=~\frac{\Gamma_q}{2}~ e^{- \Gamma_{q}t}~
[\cosh (\frac{\Delta \Gamma_q}{2} t)~+~\cos (\Delta m_q t) ]$.
\ec
Here $\Gamma_q=\frac{\Gamma_q^H~+~\Gamma_q^L}{2}$, 
$\Delta \Gamma_q=\Gamma_q^H-\Gamma_q^L$ and $\Delta m_q=m^L_q-m^H_q$, 
where $L$ and $H$ denote respectively the light and heavy physical states,
$m_q^{L(H)}$ and $\Gamma_q^{L(H)}$ are the mass and total decay width of 
these states.
The oscillation period gives a direct measurement of the mass difference between
the two physical states. 
The Standard Model predicts that $\Delta \Gamma \ll \Delta m$~\cite{ref:PDG}. 
Neglecting a possible difference between the $\rm{B^0_s}$ lifetimes
of the heavy and light mass eigenstates, the above expression simplifies to:
\bc 
${\cal P}_{B_q^0}^{unmix.}~=~{\cal P}(\mbox{B}_q^0\rightarrow \mbox{B}_q^0)~=
~\frac{1}{2 \tau_{q}} e^{- \frac{t}{\tau_{q}}}~ [ 1 + \cos ({\Delta m_q t} ) ]$
\ec
and similarly:
\bc
${\cal P}_{B_q^0}^{mix.}~=~{\cal P}(\mbox{B}_{q}^{0} \rightarrow \overline{\mbox{B}_{q}^{0}})~
=~\frac{1}{2 \tau_{q}} e^{- \frac{t}{\tau_{q}}}~ [ 1 - \cos ({\Delta m_q t} ) ].$
\ec


In the Standard Model, the $\Bq-\overline{\Bq}$ 
($q = d,s$) mixing frequency $\dmq$ (keeping 
only the dominant top quark contribution) can be expressed as follows:
\begin{equation}
\dmq = \frac{G_F^2}{6 \pi^2}~ |V_{tb}|^2~ |V_{tq}|^2~ m_t^2~ m_{B_q}~ f_{B_q}^2~ 
       B_{B_q}~ \eta_B~ F\left( \frac{m_t^2}{m_W^2} \right).
\label{eq:dmth}
\end{equation}
In this expression $G_F$ is the Fermi coupling constant; $F(x_t)$, with
$x_t=\frac{m_t^2}{m_W^2}$, results from the evaluation of the box diagram and
has a smooth dependence on $x_t$;
$\eta_B$ is a QCD correction factor obtained at next-to-leading order in 
perturbative QCD.
The dominant uncertainties in equation~(\ref{eq:dmth}) come from
the evaluation of the B-meson decay constant $f_{B_q}$ 
and of the ``bag'' parameter $B_{B_q}$.

The two elements of
the $V_{CKM}$ matrix are equal to:
\begin{equation}
 |V_{td}| = A \lambda^3 \sqrt{( 1- \rho )^2 +  \eta^2}  ~~~~~ ;~~~~~        |V_{ts}| = A \lambda^2,
\end{equation}
neglecting terms of order ${\cal O}(\lambda^4)$. $\mid V_{ts} \mid$ is independent of $\rho$ and $\eta$
and is equal to $\mid V_{cb} \mid$.\\
$\dmd$ has been measured precisely  by several experiments. Nevertheless this precision
cannot be fully exploited due to the large uncertainty
which originates in the evaluation of the non-perturbative QCD parameters.
The ratio between the Standard Model expectations for $\dmd$ and $\dms$ is given by the following expression:
\begin{equation}
\frac{\dmd}{\dms}~=~\frac{ m_{B_d}~ f^2_{B_d}~ B_{B_d}~ \eta_{B_d}}
                         { m_{B_s}~ f^2_{B_s}~ B_{B_s}~ \eta_{B_s}}~ 
	  \frac{\left | V_{td} \right |^2}{\left | V_{ts} \right |^2}.
\label{eq:ratiodms}
\end{equation}
 A measurement of the ratio $\frac{\dmd}{\dms}$ gives the same type of 
constraint in the ${\rho}-{\eta}$ plane, as a measurement of $\dmd$, 
and this ratio only depends on $f_{B_d}/f_{B_s}$ and $B_{B_d}/B_{B_s}$ 
which can be predicted with better precision than the absolute values.

Using existing measurements which constrain $\rho$ and $\eta$,
except those on $\dms$, the distribution for
the expected values of
$\dms$ can be obtained. It has been shown that $\dms$ has to lie,
at the 68$\%$ C.L., between 10.9~ps$^{-1}$ and 18.1~ps$^{-1}$ 
and is expected to be smaller than
$21.5$~ps$^{-1}$ at the 95$\%$ C.L.~\cite{ref:bello}.

\section{The inclusive lepton analysis}
\label{sec:2}

 For a description of the DELPHI detector and its performance,
the reader is referred to~\cite{ref:DELPHI}.
 The analysis described in this paper 
used precise tracking based on the silicon
microvertex detector (VD) to reconstruct the primary and secondary
vertices. To estimate the B-meson momentum and direction,   
the neutral particles detected in the electromagnetic 
and hadronic calorimeter and the reconstructed charged particle tracks were 
used. Muon identification 
was based on the hits in the muon chambers being associated with a track.
Electrons were identified using tracks associated with a shower in the 
electromagnetic calorimeter. The $dE/dx$ energy loss measurement in the 
Time Projection Chamber (TPC) and the Cherenkov light detected 
in the Ring Imaging Cherenkov detector (RICH) were used to separate pions (and also electrons or muons) 
from kaons and protons. 

Charged particles were selected if they satisfied the following criteria:
a momentum larger than 200~MeV$/c$, a track length larger than 30~cm, 
a relative momentum error smaller than 130\%, 
a polar angle ($\theta$, defined with respect to the beam axis) 
between 20$^{\circ}$ and 160$^{\circ}$
and an impact parameter with respect to the primary vertex, determined on
an event-by-event basis, smaller than 4~cm 
in the $xy$ plane (perpendicular to the beam axis) and 10~cm in $z$ (along the 
beam direction). 
Neutral particles had to deposit at least 500~MeV in the calorimeters and 
their polar angles had to lie between 2$^{\circ}$ and 178$^{\circ}$.

To select hadronic events it was required that more than~7 
charged particles were accepted with a total energy larger than 15~GeV. 
The sphericity direction was determined using charged and neutral particles and 
its polar angle was 
required to satisfy $|\cos\theta_{sphericity}|<0.8$.   
The event was divided into two hemispheres by a plane perpendicular 
to the sphericity axis. In each hemisphere
the total energy from charged and neutral particles had to be larger than 5~GeV.
A total of about 4 million hadronic Z decays were selected 
from which 3.5 million 
were taken in the LEP~I phase (1992-1995) and 0.5 million were 
collected as calibration data in the LEP~II phase (1996-2000).
Two different detector periods were analysed separately: 
1992-1993 and 1994-2000. 
In the 1992 and 1993 data the vertex detector measured only 
the $R\phi$ coordinate
($R$ being defined as $\sqrt{x^2+y^2}$ and $\phi$ is the azimuthal angle),
while from 1994 to 2000 the $z$ coordinate was also measured.

 Using tracks with vertex detector information, the primary vertex was fitted 
using the average beam spot as a constraint~\cite{ref:borissov}.

 Jets were reconstructed using charged and neutral particles by 
the LUCLUS~\cite{ref:jetset} jet algorithm 
with a transverse momentum selection cut $d_{join}$ of 6~GeV$/c$.


 Muons were identified by combining the measured position of the muon chamber 
hits with the tracking information. The tracks of charged particles were 
extrapolated to the muon chambers and then associated and fitted to the hits. 
The muon identification algorithm is described in~\cite{ref:DELPHI}.
``Loose" identified muons with momenta larger than
3~GeV$/c$ were accepted, as well as ``standard" and ``tight" identified muons 
with momenta larger than 2~GeV$/c$. 

The electron candidates were identified by combining the electromagnetic 
shower information from the High density Projection Chamber (HPC) 
with the particle ionization loss, $dE/dx$, measured by the TPC.
A sizeable fraction of electrons originates from photon conversions. 
They were partially rejected
if two oppositely charged particles formed a secondary vertex where the 
invariant mass was zero within measurement errors. 
The different variables were combined using a neural network~\cite{ref:DELPHI}.
Selection cuts on the output of the network were adjusted 
to vary with the particle momentum ensuring a constant efficiency 
between 2~GeV$/c$ and 30~GeV$/c$. 
Two levels of selection (``standard'' and ``tight'') were tuned to provide 
75$\%$ and 65$\%$ efficiency for electrons from B semileptonic decays, 
respectively. 

 Samples of  hadronic Z decays (4 million events) and of Z bosons decaying
only into $\rm b\bar{\rm b}$ quark pairs (2 million events) were 
simulated using the Monte Carlo generator JETSET~7.3~\cite{ref:jetset}
with DELPHI-tuned JETSET parameters 
and updated $b$ and $c$ decay tables~\cite{ref:tuning}. 
The detailed response of the DELPHI detector was simulated~\cite{ref:DELPHI}. 

 The principle of the oscillation measurements is as follows. 
Each of the charged and neutral particles measured in the event
is assigned to one of the two hemispheres defined by the plane transverse to the sphericity axis.
A ``production tag", correlated to the $b/\overline{b}$ sign of the initial quark at 
the production point, is defined using both hemispheres; in the hemisphere containing the lepton, the decay time of the 
B hadron is evaluated and a ``decay tag" is defined, correlated with the $\mbox{B}/\overline{\mbox{B}}$ nature of the 
decaying hadron. The lepton charge defines the ``decay tag".

\subsection{Discriminant analysis}
\label{p:disc}
Several algorithms and selections in this paper are based on discriminant 
variables.
The method used to build a discriminant variable is briefly described here.
Given two classes ($A$ and $B$) of events and $N$ variables that are able to 
distinguish, with different efficiencies, between
events from these two classes, the discriminant variable $R$ is defined as:
\begin{eqnarray*}
   R=\prod_{i=1,N} f^{A}(x_i)/f^B(x_i) 
\end{eqnarray*}
where $f^A(x_i)$ and $f^B(x_i)$ are the probability density functions 
for the variable $i$ in the samples $A$ and $B$, respectively 
(computed from simulated events).

 $R$ is  often rescaled, for practical reasons, in a finite range: 
in [-1,1] for flavour tagging ($X=(R-1)/(R+1)$) 
or in [0,1] ($X=R/(R+1)$) in all the other cases.

\subsection{Measurement of the B decay proper time}
\label{sec:21}

The B decay proper time, $t_B$, is measured from the estimates 
of the B decay distance, $d_B$, and momentum, $p_B$:
\begin{equation}
\label{eq:time}
 t_B~=~\frac{d_B ~m_B}{p_B}.
\end{equation}

\subsubsection{B decay distance}
\label{sec:211}

The B decay distance has been obtained from the measurement of the
distance between the secondary and the primary vertices, projected
along the jet direction in the $x-y$ plane and evaluated along the jet
direction.

 All charged particles with an impact parameter smaller than 2~mm relative to the
beam interaction point in the plane transverse to the beam direction
have been used to reconstruct the primary vertex. The position of the
beam spot has been used as a constraint. If the primary vertex fit had a
$\chi^{2}$-probability smaller than $10^{-3}$, an iterative procedure was
applied which removed the track contributing most to the
$\chi^{2}$ at each iteration. In a simulated  $b\overline{b}$ sample,
this procedure has been found to reconstruct vertices with an accuracy 
of 80~${\mu}$m in the horizontal $x$
direction, where  the beam spot has the larger spread, 
and 40~${\mu}$m in the vertical $y$ direction.
 
 The B decay distance was evaluated using two different algorithms. 

\begin{itemize}
\item {Mini-jets algorithm. \\ }
The position of the B decay vertex and the B momentum are obtained by using an
algorithm especially conceived for B hadron semileptonic decays.
This algorithm is based on a classification of the charged particles 
in the event as B decay products or as emitted from the beam interaction point. 
The B secondary vertex is obtained by intersecting the trajectories of the
lepton and of a D candidate. The lepton track and at least one of the charged 
particles assigned as a D
decay product have to be associated to hits in the VD.
Particles from fragmentation and from B decay products are all present in the
jet which contains the lepton, so an approach has been developed to distinguish
between them. Ignoring the lepton, charged
particles belonging to the jet are gathered into low-mass clusters, 
using LUCLUS with $d_{join}$ reduced to 0.5~\GeV$/c$ 
and assuming that the particles are pions. Inside each cluster, the particles
are ordered in decreasing values of their pseudo-rapidity relative 
to the cluster direction.
Those having the largest pseudo-rapidity values and a momentum larger 
than 500~MeV$/c$
are then kept until the mass of the resulting system exceeds 2.2~\GeV$/c^2$.
Clusters which make an angle larger than 500~mrad relative to the jet direction 
are discarded.
If a cluster contains more than one particle measured in the VD, 
a secondary vertex is obtained
from the particles belonging to the cluster, a pseudo-D track candidate is
constructed and the intersection of the pseudo-D track with the lepton 
trajectory is evaluated.
If a cluster contains only one particle measured in the VD, its intersection
with the lepton trajectory is evaluated. Among all these secondary vertices,
 the one which has the largest significance is kept~\footnote{The significance 
is defined as the distance between the primary and the secondary vertices along 
the jet direction, evaluated in the plane transverse to the beam axis, 
divided by its measurement error.}.
Having selected the cluster which contains a majority of D decay products 
and to reduce possible biases induced by this selection
on the measured decay length of the B hadron, 
this cluster is used simply as a seed to find the
other particles emitted by the D, which may be classified in other clusters. 
For this purpose,
all particles present in the jet, including neutrals but not the lepton, 
are ordered by decreasing values of their pseudo-rapidity relative to the 
direction of the momentum sum of the previously retained particles. 
Particles are then added to the previously retained ones until the mass of the
system exceeds 2.2~\GeV$/c^2$. A new evaluation of the D candidate trajectory 
is then obtained
and a secondary vertex is constructed with the lepton track. 
All of the retained particles are then called B decay products.
Once the set of tracks which contains a majority of D decay products 
is selected, the B secondary vertex is obtained by intersecting the
trajectories of the lepton and of the D candidate.

\vskip 0.2 cm
\item {Grid Algorithm.} \\
The idea is based on constructing a 3-dimensional grid, where the points 
represent all possible secondary vertices. 
This grid is built up around the B direction. 
The B direction is defined by the sphericity axis for two-jet events 
or by the axis of the jet closest to the lepton candidate. 
The resolution on the B direction 
is around 50~mrad in $\theta$ and 60~mrad in $\phi$.  
All the charged-particle tracks, other than the lepton, 
are then assigned to the primary or to any possible secondary vertex 
(any point of the grid) according to a probability which depends
on the impact parameters in the R-$\phi$ plane and along $z$ 
(computed with respect to this candidate vertex), 
the momentum and the rapidity of the particle with respect to the jet axis. 
The particles are clustered as tracks belonging to the primary vertex and 
to the candidate secondary vertex. The overall probability of each configuration 
is then computed. 
The best configuration is the one with the maximum probability and it defines
primary and secondary tracks. The D vertex is computed using the 
tracks that have been assigned to the secondary vertex.
A B vertex is then obtained
by fitting the trajectories of the lepton, the D candidate and
the B direction to a common vertex.\\
If this procedure fails, the evaluation of the B secondary vertex is performed 
by intersecting the B direction with the lepton track.
\end{itemize}

A discriminant analysis is then performed to choose the best B secondary vertex, trying to separate the secondary vertices
having a resolution better or worse than 250~$\mu$m. 
For each algorithm a discriminant variable is constructed (following the procedure
explained in \ref{p:disc}) using the error on the reconstructed decay length for the B 
and the D candidate (if any), the $\chi^2$ of the B and the D (if any) 
secondary vertex as well as the number of
VD hits for the tracks forming the B vertex.
Vertices characterized by a high (low) value of this discriminant variable
have a good (poor) resolution.

The algorithm with the largest value of the discriminant variable is chosen 
for the given hemisphere.

In order to clean up the sample from vertex candidates with poor resolution,
hemispheres with a value of the discriminant variable smaller than 0.4 
are rejected. The selection cut has been chosen, studying simulated events, 
as the highest value 
that allows the sensitivity of the $\dms$ analysis
to be kept almost unchanged.

The remaining sample is divided into five classes according to 
the value of the discriminant variable.

\subsubsection{B momentum}
\label{sec:212}

The B momentum is also evaluated using two different algorithms. \\
In the first algorithm, the B momentum is determined in several steps 
using the mini-jets method. First, each event is divided into two 
hemispheres separated by the plane transverse to the sphericity axis 
which contains the beam interaction point. 
Then the four-momentum of the B-meson, $P^B_{meas.}$, is evaluated 
by subtracting from the four-momentum of the hemisphere the sum of the four-momentta
of the particles not selected as B decay products. 
Then, to have a better estimate of the B momentum, 
the measured energies and momenta are rescaled by a common factor ($\alpha$) 
and a missing four-momentum corresponding to a zero mass particle is added 
($P_{\nu}, \overrightarrow{P_{\nu}}$).
Energy and momentum conservation, applied to the complete event, 
determine these unknowns:
\begin{eqnarray}
\alpha \; (\overrightarrow{P_{hem1}} + \overrightarrow{P_{hem2}}) +
\overrightarrow{P_{\nu}} & = & \overrightarrow{0} \\
\alpha \; ( E_{hem1} + E_{hem2}) + P_{\nu} & = & 2~ E_{beam}.
\end{eqnarray}
The mean value of $\alpha$, determined using simulation, is 1.13.
If the direction of the missing momentum lies within 400~mrad of the direction 
of the D-$\ell$ system, its energy is
attributed to the B to account for the missing neutrino. 
A better approximation to the B momentum is then obtained using
the simulation, by correcting for the average difference between the above 
estimator and the true B momentum, parameterised
as a function of the reconstructed B momentum.
Finally a global fit is applied to all the measured quantities: the primary and secondary vertex positions (6 variables),
and the momentum vectors of the lepton and of  the D and B-mesons (9 variables). Three constraints are applied:
\begin{itemize}
\item the direction given by the two vertices and the direction of the B 
momentum should be the same (two angular constraints),
\item the mass of the B-meson should be equal to the nominal $\Bs$ mass.
\end{itemize}

The second algorithm evaluated the B energy using the 
BSAURUS package~\cite{ref:lifebuda}.
A discriminant analysis is then performed to choose the best B energy, 
trying to discriminate the candidates having a relative
energy resolution better than 8$\%$.

The choice between the two algorithms is made using the same method as the one 
described for the vertex resolution.

In order to clean up the sample from B candidates with poor momentum 
resolution,
hemispheres with a value of the discriminant variable smaller than 0.3 
are rejected. This value has been chosen using the same procedure as the one 
described for the vertex resolution.

The remaining sample is divided into three classes according to the value 
of the discriminant variable. 

\subsubsection{Parameterisation of the resolution functions}
\label{sec:22}

To have a detailed description of the time resolution, 
the distribution of the difference between the generated ($t_{sim}$) and
reconstructed ($t_{rec}$) B decay proper times ${\cal R}_B(t_{sim}-t_{rec})$ 
has been parameterised by using the sum of Gaussian distributions with
widths that depend on the generated decay time and 
on the sign of $t_{sim}-t_{rec}$.
For about 10$\%$ of the events, because of the presence of 
charged-particle tracks coming from the beam interaction point,
the reconstructed vertex co{i}ncides with the event main vertex. 
A Gaussian distribution, centred on $t_{rec}=0$, has been used to account for 
the proper time distribution of these events.

The accuracy of the measurement of the B decay proper time is degraded for 
cascade semileptonic decays ($b \rightarrow c \rightarrow \ell X$), 
since the parameterisation of the difference 
between the true and measured B momentum and the strategy
for the reconstruction of the B decay point have been tuned assuming a direct 
semileptonic decay of a B  hadron.
This has been verified using the simulation and a different parameterisation 
${\cal R}_{BC}(t_{sim}-t_{rec})$ has been obtained for this category of leptons.

The function used to parameterise the resolution is: 
\begin{itemize}
\item for the central part: three asymmetric Gaussian distributions~\footnote{
An asymmetric Gaussian distribution consists of two Gaussian distributions  
with the same central value and amplitude but with different widths
in the region $x<\mu$ and $x>\mu$.}
with identical central values.
The parameterisation contains 10 parameters: 3 fractions $f_{i=1,3}$ ($f_1>f_2>f_3$), 1 central value ($x_1 = x_2 = x_3 = \mu$) and 6 widths 
${\sigma}_i,{\sigma}_i'$.
\begin{eqnarray}
  \label{eqgadef}
  G^{(x_i=\mu , \sigma_i, \sigma^{\prime}_i)}(x) & \equiv & \frac{1}{\sqrt{2\pi}} \frac{2}{\sigma_i+\sigma^{\prime}_{i}} e^{-\frac{(x-\mu)^2}{2{\sigma_i}^2}},
x \le \mu \nonumber \\
 & \equiv & \frac{1}{\sqrt{2\pi}} \frac{2}{\sigma_i+\sigma^{\prime}_i} e^{-\frac{(x-\mu)^2}{2{\sigma^{\prime}_i}^2}}, x \ge \mu
\end{eqnarray}

\item for the primary vertex, a Gaussian distribution corresponding to the fraction $f_4$, of central value $x_4$ and width 
$\sigma_4$.
\end{itemize}

The general expression for the time resolution is:
\begin{equation}
  \label{eqres}
{R}(t_{sim}-t_{rec},t_{sim}) = \sum_{i =1,4} f_i~G^{(x_i(t_{sim}),{\sigma}_i(t_{sim}),r_i(t_{sim}))}(t_{sim}-t_{rec})
\end{equation}
where $r_i \equiv \frac{{\sigma}_i}{{\sigma}^{\prime}_i}$.
The fractions $f_i$ are normalised in the following way:
\begin{equation}
  \label{eqnorm}
  f_4=1-f_1-f_2-f_3.
\end{equation}
This parameterisation contains 12 free parameters
which depend on the generated time. 
The time dependence of these parameters is given in Table~\ref{tab:tabpart}.
The 29 quantities from $a$ to $m$ (see Table~\ref{tab:tabpart}) are determined 
by fitting the $t_{sim}-t_{rec}$ 
distribution from the simulation as a function of $t_{sim}$. 
Finally the simulated time distribution for accepted events is compared with an 
exponential distribution corresponding to the generated lifetime and an acceptance 
function, ${\cal A}(t_{sim})$, is obtained. 
In practice ${\cal A}$ is constant over the
accepted range of decay times between 0 and 12 ps.
For events originating from light and charm quark flavours, 
the expected time distributions, ${\cal P}_{H}(t)$ and 
${\cal P}_{C}(t)$, have been obtained from the simulation. 
The distributions are parameterised using a combination of three 
exponentials convoluted with Gaussian distributions.

\begin{center}
\begin{table}[ht]
\begin{center}
\begin{tabular}{|c|} \hline
 $f_{1,2,3} = a_{1,2,3} + b_{1,2,3} \cdot t_{exp} + c_{1,2,3} \cdot t_{exp}^2$ \\ \hline
 $f_4 = 1-f_1-f_2-f_3$ \\ \hline
 $\mu = d+e \cdot t_{exp} + f \cdot t_{exp}^2 + g \cdot t_{exp}^3$ \\ \hline
 $x_1 = x_2 = x_3 = \mu$ \\ \hline
 $x_4 = h + i \cdot t$ \\ \hline
 ${\sigma}_{1,2,3,4} = \sqrt{j_{1,2,3,4}^2 + k_{1,2,3,4}^2 \cdot t_{exp}^2}$ \\ \hline
 $r_{1,2,3} = l_{1,2,3} + m_{1,2,3} \cdot t_{exp}$ \\ \hline
 $r_4 = 1$ \\ \hline
\end{tabular}
\end{center}
\caption{ \it {Time dependence of the parameters entering into the 
parameterisation of the resolution function, where $t_{exp} 
\equiv 1-exp(\frac{-t}{\tau})$.} }
\label{tab:tabpart}
\end{table}
\end{center}

\begin{figure}[htb]
\begin{center}
\epsfig{figure=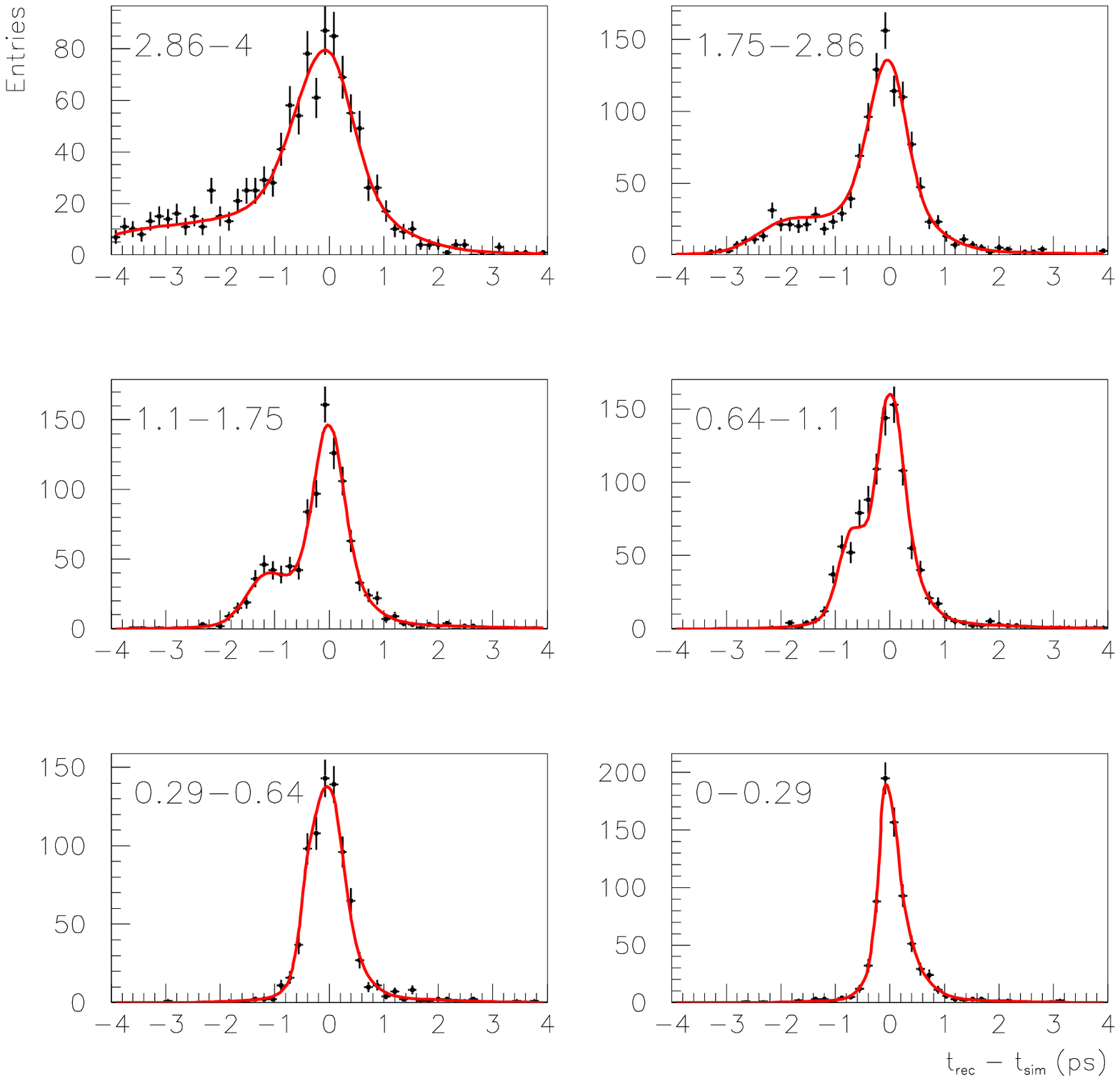,height=16cm}
\caption{  \it  {The distributions of $t_{rec}-t_{sim}$ in different intervals 
of $t_{sim}$ as indicated in top right corner of each plot (the unit is ps), 
for the worst class in proper-time resolution on simulated 1994-2000 data
(points with error bars).
The satellite peak appearing on the left part of the distribution corresponds 
to events in which the primary vertex has
been reconstructed instead of the real B decay vertex.
The curves correspond to the parameterisation explained in
section~\ref{sec:22}.}}
\label{fig:plotrisoluzione_cl1}
\end{center}
\end{figure}

\begin{figure}[htb]
\begin{center}
\epsfig{figure=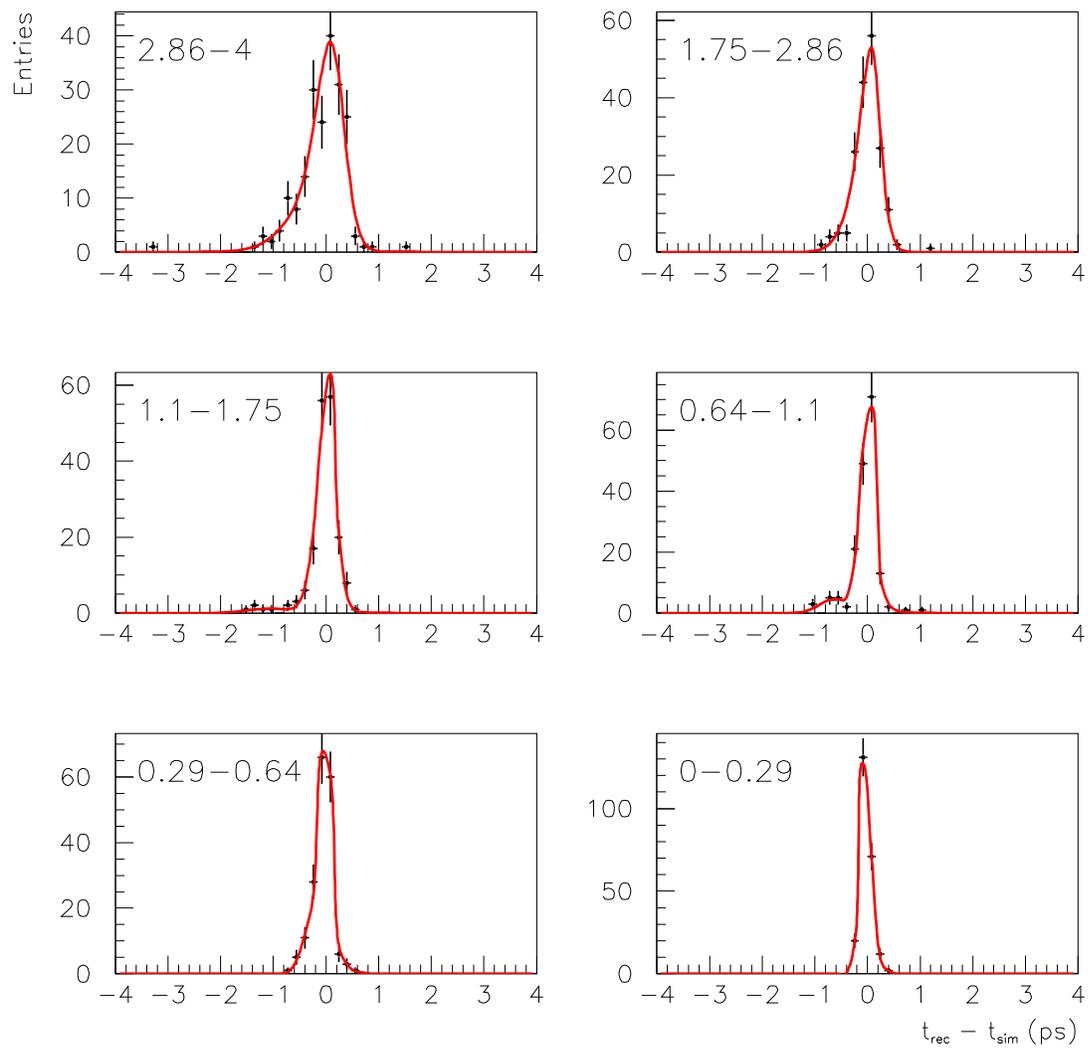,height=16cm}
\caption{  \it  {Same as in Figure~\ref{fig:plotrisoluzione_cl1} 
for the best class in proper-time resolution.}}
\label{fig:plotrisoluzione_cl15}
\end{center}
\end{figure}

Finally the sample was uniformely divided into 15 subsamples according to 5 classes 
in decay-length resolution and 3 classes in momentum resolution.
Figures~\ref{fig:plotrisoluzione_cl1} and~\ref{fig:plotrisoluzione_cl15} show, 
as an example, the distributions of 
($t_{rec}-t_{sim}$) in different intervals of $t_{sim}$ as indicated 
in each histogram for the events which have respectively the worst and the best
proper time resolution for the 1994-2000 data.

\subsection{Algorithm for tagging the \boldmath $b$ flavour at production time}

The signature of the initial production of a $b \; (\overline{b})$ quark
in the jet containing the $\Bs$ or $\overline{\Bs}$ candidate
is determined in the opposite hemisphere, from a combination of different 
variables sensitive to the initial quark state. 
These variables are combined in a devoted neural network 
(BSAURUS~\cite{ref:lifebuda}).
The output of this neural network, $x^{opp.}_{tag}$, varies between -1 and 1. 
Values of $x^{opp.}_{tag}$ close to 1 (-1) correspond to a high probability 
that a given hemisphere contains a $b$ ($\bar b$) quark in the initial state. 
This tagging is complemented in the same hemisphere with a simple discriminant 
variable using the information from tracks which are not attached to the 
reconstructed B vertex. Two variables are used:
\begin{itemize}
\item
the mean jet charge which is defined as:
\begin{equation}
\label{eq2}
 Q_{hem}= \frac{\sum_{i=1}^{n}q_i (\vec{p_i} \cdot \vec{e_s})^{\kappa}}
{\sum_{i=1}^{n} (\vec{p_i} \cdot \vec{e_s})^{\kappa}}.
\end{equation}
In this expression, $q_i$ and $\vec{p_i}$ are respectively the charge and the 
momentum of particles $i$, $\vec{e_s}$ is the unit vector along the sphericity 
axis and $\kappa$=0.6;
\item 
the rapidity of the charged particle with highest rapidity among
the identified kaons in the same hemisphere (spectator kaon).
\end{itemize}
The output of the neural network in the opposite hemisphere, 
the mean jet charge and rapidity
of the kaon in the same hemisphere are combined into
a single discriminant variable $x_{tag}$ as described in section~\ref{p:disc}.

The distributions of $x^{oppo}_{tag}$ and $x_{tag}$ for simulation and data
will be presented in sections~\ref{sec:261} and~\ref{p:dms_syst}.

\subsection{Composition of the lepton sample}
\label{sec:23}

 

The muons were selected in the ``tight'' and ``standard'' categories 
while the electrons were required to belong to the  ``tight'' category.

The efficiency to identify leptons and the hadronic contamination have been
obtained by using the detailed simulation code
of the DELPHI detector~\cite{ref:DELPHI} and have been checked on data by using 
selected event samples
such as $\mbox{K}^0_s\rightarrow \pi^+\pi^-$, $\Zz\rightarrow\mu^+\mu^-$, 
photons converted
in front of the HPC, $\gamma \gamma \rightarrow \ell^+ \ell^-$ 
and hadronic $\tau$ decays~\cite{ref:DELPHI}.

Candidate leptons are produced by semileptonic decays of B hadrons, D hadrons,
and light mesons. They also can be misidentified hadrons or converted photons. 
Leptons from cascade decays have the wrong sign with 
respect to leptons from direct B decays for the identification of the $b$ quark 
charge. 
Thus, selections have been defined on the lepton total, $p$, 
and transverse, $p_t$, momenta to minimize their contribution\footnote{These 
selections have been obtained by maximizing 
the product $(f_b^l-f_{bc}^l) \times \sqrt{\mbox{N}_l}$, 
where $f_b^l$ and $f_{bc}^l$ are the fractions of direct and
cascade leptons in the total sample of N$_l$ lepton candidates.}.
The fractions of the different categories of selected leptons, 
with $p$ larger than 3~GeV$/c$ and $p_t$ larger than 1.2~GeV$/c$, 
have been measured using the real data by fitting the ($p, p_t$) distributions 
of the different components in the simulation 
to the corresponding distribution for lepton candidates 
observed in the data.
The transverse momentum is computed with respect to the jet axis after having 
removed the lepton from the jet.

To study $\mbox{B}-\bar{\mbox{B}}$ oscillations, 
lepton candidates in the simulated event sample have been divided into four 
classes according to their sign relative to the sign of the heavy quark present 
in the decaying hadron:
 
\begin{itemize}
\item
$\ell_{b}$: lepton candidates, produced in B hadron decays, having
the same sign as the $b$ quark present inside the B hadron. This class contains
leptons from direct
$b$ semileptonic decays including $\tau$ decay products and also, for example, those from cascade decays of the
type: $\overline{\mbox{B}} \rightarrow \mbox{D} \overline{\mbox{D}} X$ with 
$\overline{\mbox{D}} \rightarrow \ell^- X$. Misidentified hadrons with the 
same sign as the $b$ quark are also included if they originate from a 
B hadron decay. 
 
\item
$\ell_{bc}$: when the candidate lepton has a sign opposite to the $b$ quark
charge.
 
\item
$\ell_{c}$: when the candidate lepton is produced in a charmed hadron decay which is not coming from a B-meson decay.
 
\item
$\ell_{h}$: for candidate leptons which are misidentified hadrons
or leptons produced in light hadron decays or converted photons and which
do not fall into the previous classes.
\end{itemize}
 
With these definitions, if a B hadron oscillates, lepton candidates in the 
$\ell_b$ and $\ell_{bc}$ classes change their sign, 
whereas those belonging to the other classes are not affected.

The semileptonic decay branching fractions used in the simulation have been 
modified to agree with the measured values. 
The fraction of fake lepton candidates has been verified in data with an 
accuracy of $\pm 10\%$~\cite{ref:margoni}. 
The Monte Carlo simulation is then corrected so that the rates of the physics processes and 
their simulation agree with present  measurements. 
As an example, for 1994-2000 data, the fractions of lepton candidates 
in the worst and best classes in proper time resolution are:
\begin{eqnarray}
   \begin{array}{lll}
f_{b}^l=67.6 \%,~f_{bc}^l=15.4 \%,~f_{c}^l=4.8 \%, ~f_{h}^l=12.1\%~(1994-2000~~{\mathrm worst~class}) & \\
f_{b}^l=87.2 \%,~f_{bc}^l=7.7 \%,~~f_{c}^l=3.2 \%, ~f_{h}^l=~1.9\%~(1994-2000~~{\mathrm best~class}). & \\
   \end{array}
 \end{eqnarray}

Uncertainties attached to these fractions depend on those on the semileptonic
branching fraction measurements, on the lepton identification procedure and
on the rate of the fake lepton background. 
They will be discussed in section~\ref{res}.

\begin{figure}[htb]
\epsfig{figure=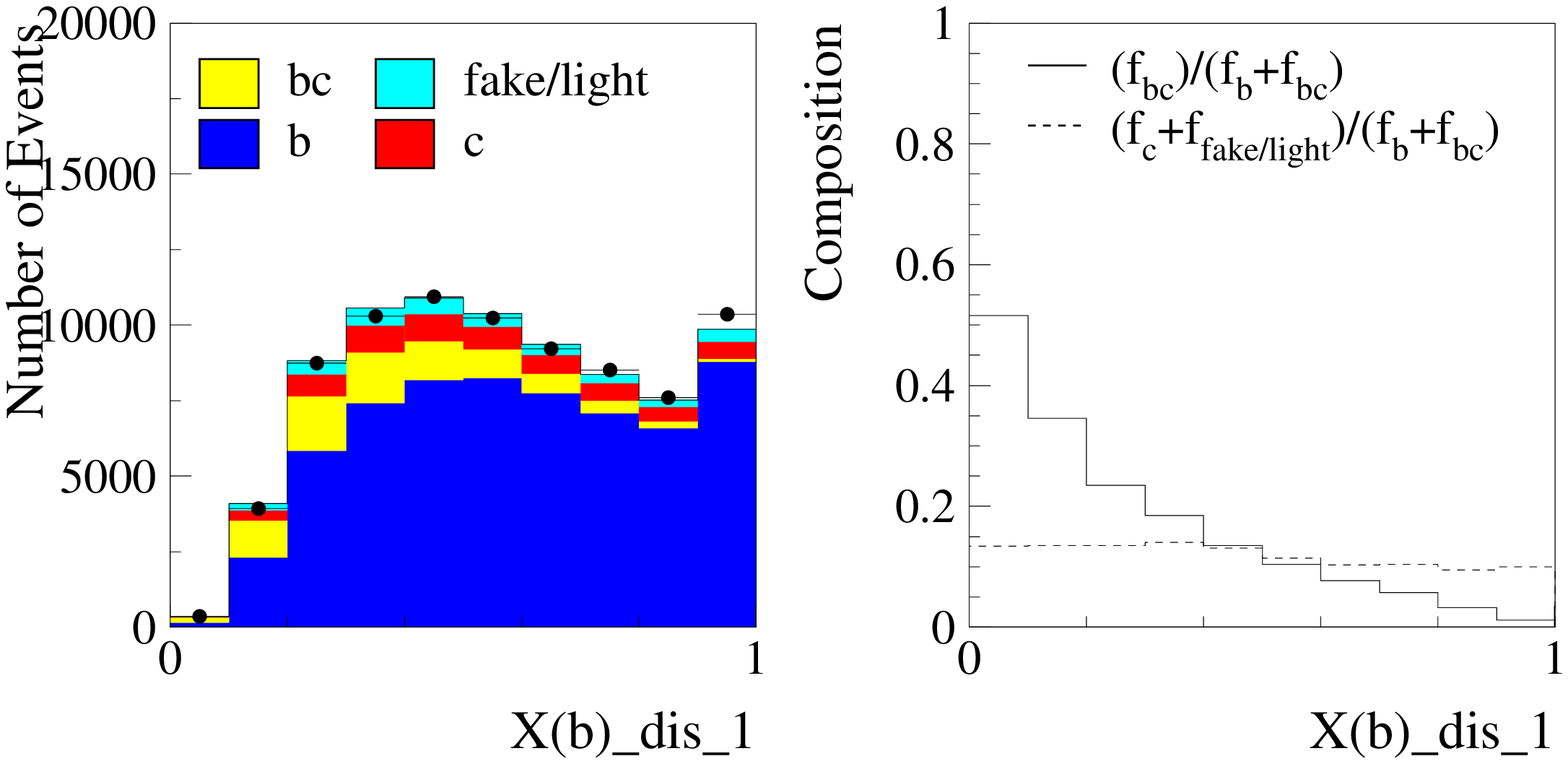,height=8cm}
\epsfig{figure=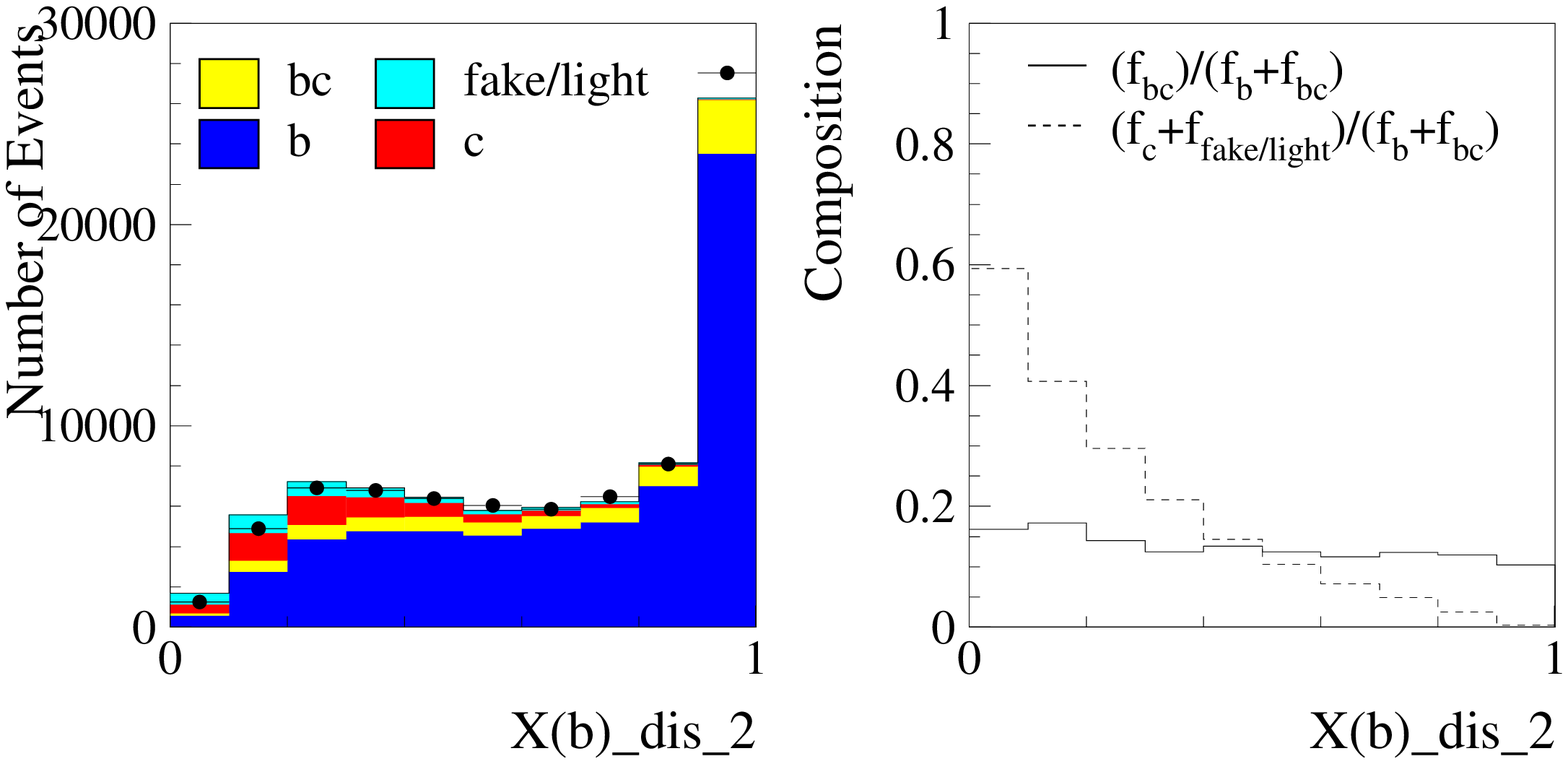,height=8cm}
\caption{ \it The plots shows the distribution of the discriminant variables 
used to distinguish between the events 
in which the lepton candidate is coming from direct B decays ($bl$), 
cascade decays ($bcl$), charm decays ($c$) and 
misidentified hadrons ($fake$). 
They are shown for 1994-2000 data (full dots).
The plots on the top (bottom) show the projection onto the discriminant 
variable used to distinguish between events in which the leptons come from 
direct and cascade B decays (from B decays ($bl$ or $bcl$) and the rest).
The data/Monte Carlo agreement is shown on the left.
The figures on the right show the composition of the simulated sample 
as a function of the value of the discriminating variable.}
\label{fig:discrbl_94}
\end{figure}

To improve the separation between the class corresponding to the signal, 
$\ell_b$, and the other classes, a set of six variables has been used:
\begin{itemize}
\item $p$, the momentum of the reconstructed lepton;
\item $p_t$, the transverse momentum of the reconstructed lepton 
with respect to the jet axis after having removed the lepton from the jet;
\item $y_{b-tag}$, the $b$ tagging variable. 
This variable is defined as the probability for the track impact 
parameters relative to the beam interaction position, 
in the hemisphere opposite to the one containing the lepton,
to be compatible with the one expected for light quark 
events~\cite{ref:borissov}. Small values for this variable tag events 
containing b-hadrons;
\item the flight distance between the $\rm B$ and the $\rm D$
mesons;
\item the multiplicity at the secondary vertex;
\item the lepton identification.
\end{itemize}
Two different discriminant variables were constructed 
(see section~\ref{p:disc}).
The first variable distinguishes between events in which the lepton candidate 
is coming from B decays (direct and cascade), 
D decays or is a misidentified hadron. 
The second one differentiates between leptons from direct and cascade B decays. 
The first discriminant variable uses all the above mentioned variables, 
while in constructing the other, the 
$y_{b-tag}$ and the lepton identification were removed. 
Two-dimensional distributions of the discriminant variables 
were then constructed for each of the proper-time resolution classes. 
Figure~\ref{fig:discrbl_94} shows the projections onto these discriminant 
variables. The fractions of the different components in the sample as a 
function of the two variables are also shown in 
Figure~\ref{fig:discrbl_94}. 
This information will be used on an event-by-event basis.

\subsection{The fraction of B hadrons of flavour \boldmath $q$ in the sample: $f_{B_q}$}
\label{sec:24}

Besides other inputs like product branching fractions involving 
characteristic signatures, precise information on 
$f_{B_s}$ can  be obtained by comparing
the integrated oscillation rates of neutral B-mesons ($\overline{\chi}$), measured at LEP and SLD,
and of $\Bdb$ mesons only ($\chi_d$), measured at LEP and at the $\Upsilon$(4S):
\begin{equation}
\overline{\chi}= f_{B_d} \chi_d + f_{B_s} \chi_s
\end{equation}
together with the normalisation condition and using isospin symmetry in the non-strange B-meson sector:
\begin{equation}
 f_{B_d} + f_{B^+} + f_{B_s} + f_{b-baryon}~=1 ~~ , ~~ f_{B^+} = f_{B_d} .
\end{equation}
Recent determinations of these fractions are reported in Table~\ref{tab:val}.
\begin{table}
\begin{center}
\begin{tabular}{|l|l|}
 \hline
  Parameter & Value \\ \hline \hline
  $f_{B^+}$, $f_{B_d}$      & $0.388\pm0.013$ \\ \hline
  $f_{B_s}$                 & $0.106\pm0.013$ \\ \hline
  $f_{\mbox{$b$-baryons}}$  & $0.118\pm0.020$ \\ \hline
  $\tau_{B^+}$                  &  $1.674\pm0.018$~ps \\ \hline
  $\tau_{B_d}$                  &  $1.542\pm0.016$~ps \\ \hline
  $\tau_{B_s}$                  &  $1.461\pm0.057$~ps \\ \hline
  $\tau_{\mbox{$b$-baryons}}$   &  $1.208\pm0.051$~ps \\ \hline
  $\dmd$           & $0.489\pm0.008$~ps$^{-1}$ \\ \hline
\end{tabular}
\end{center}
\caption{List of the values used for the most relevant 
parameters~\cite{ref:PDG}.}
\label{tab:val}
\end{table}

\begin{figure}[htb]
\begin{center}
\epsfig{figure=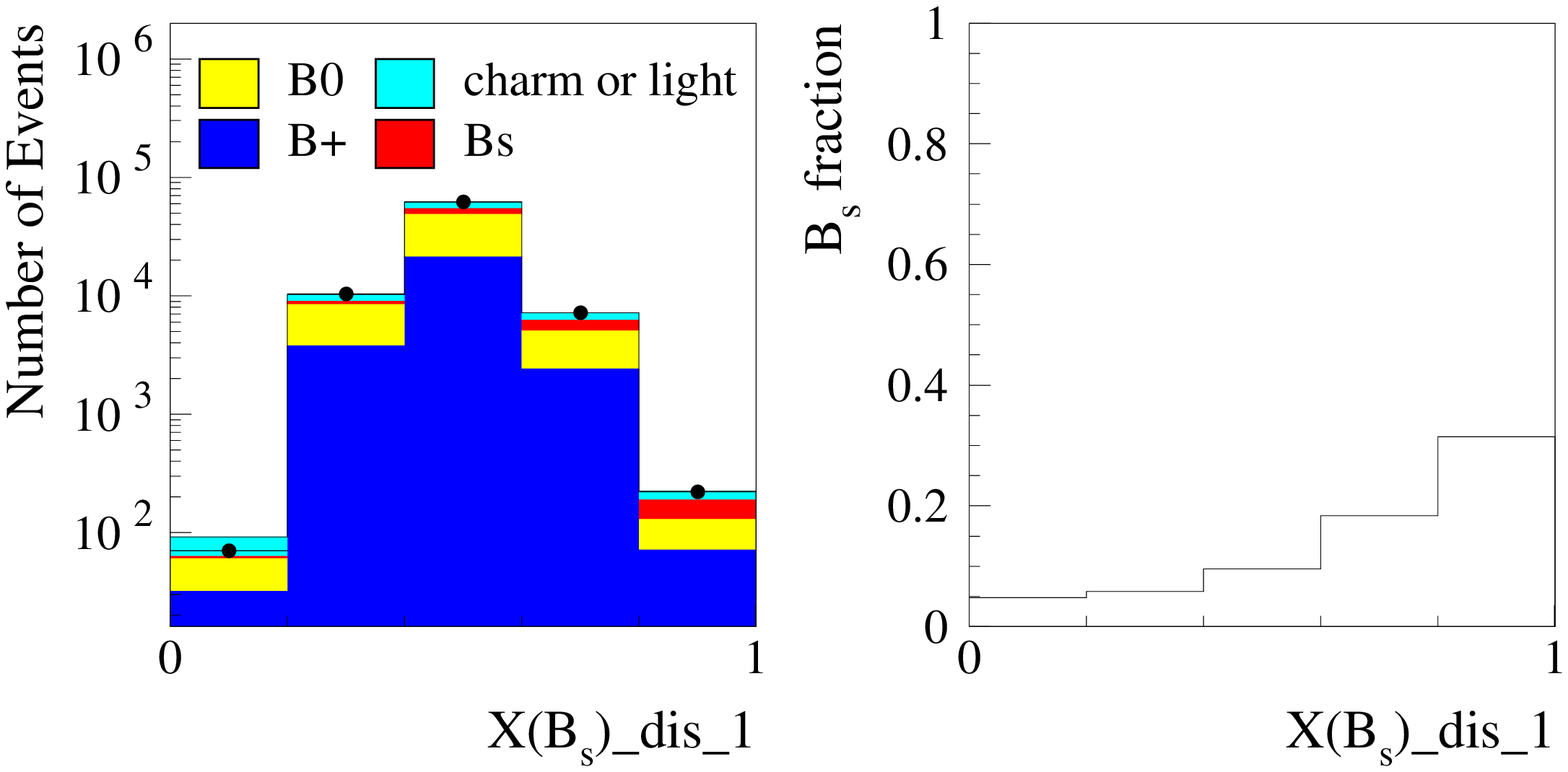,height=8cm} 
\epsfig{figure=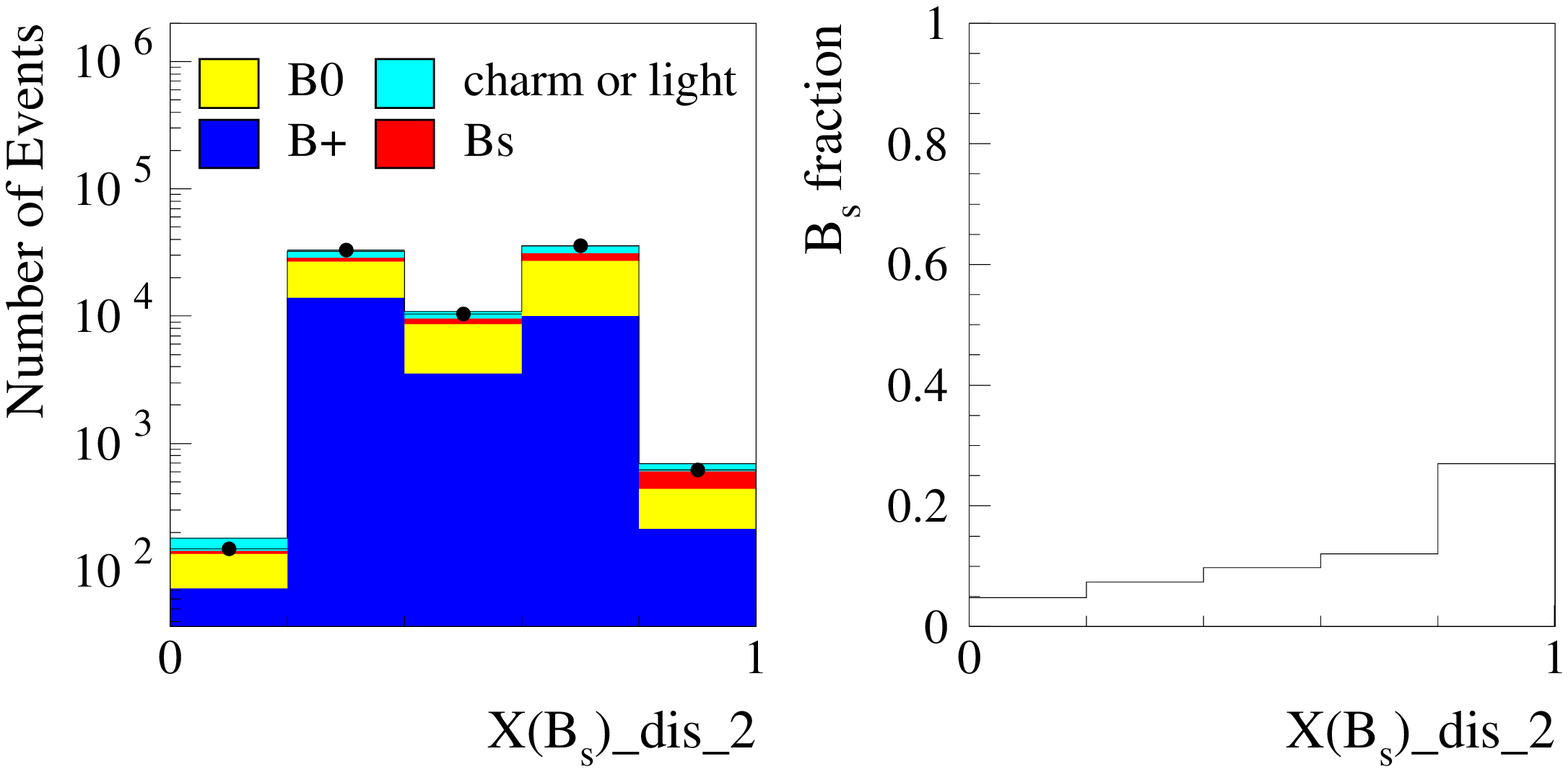,height=8cm}
\caption{\it The plots show the distribution of the discriminant variables used 
to distinguish between the events coming from
$\Bs$ decays from those coming from other neutral b-hadron decays (top) 
and from $B^+$ decays (bottom). 
They are shown for 1994-2000 data (full dots).
On the left the data/Monte Carlo agreement can be appreciated, 
on the right the figures show the $\Bs$ fraction as a function of the value of 
the discriminating variable.}
\label{fig:discrbs_94}
\end{center}
\end{figure}

To increase the fraction of $\Bs$ mesons in the sample, a set of four variables 
which are sensitive to the presence of strange B-mesons in a jet is defined:
\begin{itemize}
\item the number of kaons associated to the secondary vertex; 
\item the number of kaons coming from the primary vertex; 
\item the multiplicity of charged-particle tracks at the secondary vertex; 
\item the charge of the reconstructed secondary vertex. 
\end{itemize}
Probability distributions of these variables have been combined to define
two discriminant variables.
Only the first three variables were used to discriminate between $\Bs$ and the 
other neutral hadrons, 
while all the four variables were used to discriminate between $\Bs$ and 
charged B-mesons.
Two-dimensional distributions of the discriminant variables (see
section~\ref{p:disc}) were then constructed for each of the resolution classes. 
Figure~\ref{fig:discrbs_94} shows the projections onto the two discriminant 
variables for the 1994-2000 data, together with the variation of 
the $\Bs$ fraction as a function of the two discriminating variables. 
This information will be used on an event-by-event basis.

\subsection{Fitting procedure}
\label{sec:25}

 Events have been classified on the basis of the charge of the lepton, 
$\mbox{Q}_{\ell}$, and $x_{tag}$. 
They are considered as mixed if $x_{tag} \times \mbox{Q}_{\ell} < 0$ and as 
unmixed if $x_{tag} \times \mbox{Q}_{\ell} > 0$.

The number of events are 16448 (10160) and 26132 (15368) in the mixed and 
unmixed categories, respectively, for the 1994-2000 (1992-1993) data. 
So 68108 events have been used for the analysis. 

In each event the probability to obtain, for a given measured proper time $t$, 
a like-sign pair (charge of the lepton equal to the charge of the quark at the 
production time: ${\cal P}^{like}(t)$)
or an unlike-sign pair (charge of the lepton opposite to the charge of the 
quark at the production time: ${\cal P}^{unlike}(t)$) has been evaluated:  

\begin{eqnarray}
\begin{array}{ll}
{\cal P}^{like}(t) & =~f_b^{\ell}  \sum_{q}~f^{b}_{B_q} \epsb 
 ({\cal P}_{B_q}^{mix.}(t')
 \otimes {\cal R}_{B}(t'-t))\\
  &+~~f_{bc}^{\ell}  \sum_{q}~f^{bc}_{B_q}(1-\epsb)
  ({\cal P}_{B_q}^{mix.}(t')
 \otimes {\cal R}_{BC}(t'-t))\\
  & +~~f_b^{\ell}  \sum_{q}~f^{b}_{B_q}(1-\epsb)
  ({\cal P}_{B_q}^{unmix.}(t')
 \otimes {\cal R}_{B}(t'-t))\\
  & +~~f_{bc}^{\ell} \sum_{q}~f^{bc}_{B_q}\epsb
  ({\cal P}_{B_q}^{unmix.}(t')
 \otimes {\cal R}_{BC}(t'-t))\\
  &  +~~f^{\ell}_c  \epsilon_c^{like} {\cal P}_{C}(t) 
 + f^{\ell}_h  \epsilon_h^{like} {\cal P}_{H}(t),
\end{array}
\label{eq:ar1}
\end{eqnarray}
 
\begin{eqnarray}
\begin{array}{ll}
{\cal P}_{unlike}(t) & =~f_b^{\ell}  \sum_{q}~f^{b}_{B_q}\epsb  
  ({\cal P}_{B_q}^{unmix.}(t')
 \otimes {\cal R}_{B}(t'-t))\\
  &+~~f_{bc}^{\ell}  \sum_{q} ~f^{bc}_{B_q}(1-\epsb)
 ({\cal P}_{B_q}^{unmix.}(t')
 \otimes {\cal R}_{BC}(t'-t))\\
  & +~~f_b^{\ell}  \sum_{q}~f^{b}_{B_q}(1-\epsb)
 ({\cal P}_{B_q}^{mix.}(t')
 \otimes {\cal R}_{B}(t'-t))\\
  & +~~f_{bc}^{\ell} \sum_{q}~f^{bc}_{B_q}\epsb
 ({\cal P}_{B_q}^{mix.}(t')
 \otimes {\cal R}_{BC}(t'-t))\\
  & +~~f^{\ell}_c  (1 -\epsilon_c^{like}) {\cal P}_{C}(t)
+f^{\ell}_h  (1 -\epsilon_h^{like}) {\cal P}_{H}(t).
\end{array}
\label{eq:ar2}
 \end{eqnarray}
where the generated proper time is denoted $t'$; 
$f_b^{\ell}$, $f_{bc}^{\ell}$, $f_c^{\ell}$, $f_h^{\ell}$ are the fractions 
defined on an event-by-event basis as explained in section~\ref{sec:24}; 
$\epsilon_b^{tag}$ is the tagging purity at production time 
(which is the fraction of events where the charge of the $b$ at the production
time is correctly assigned). 
In the fit procedure, the p.d.f. 
(probability density functions) to have the right (wrong) sign for the quark at 
production and decay times are directly used.
To cope with possible differences between data and simulation, 
for the opposite hemisphere tag, the p.d.f. of $x^{oppo}_{tag}$
determined in the $\dmd$ fit (see section~\ref{sec:261}) is used.\\
The p.d.f. for the same-hemisphere tagging variables are taken from simulated data.\\
The opposite and same-hemisphere tagging variables are combined
in one single tagging variable $x_{tag}$.

In $\Zz \rightarrow b \overline{b}$ events, $f^{b}_{B_q}$ and $f^{bc}_{B_q}$
are the fractions of direct and cascade decays, respectively, of B hadrons 
of flavour $q$ in the sample. 
For direct decays it has been assumed that these fractions
are the same as the corresponding production rates of the different B hadrons 
in $b$ jets because very
similar semileptonic partial widths are expected for all B hadrons
(they are equal for $\Bdb$ and $\Bm$ due to isospin invariance). 
The cascade decay fractions have been computed from Monte Carlo simulation
after correcting the production rates.

In $\Zz \rightarrow c \overline{c}$ events and for light flavours,
$\epsilon_c^{like}$ and $\epsilon_h^{like}$ are the fractions of events
classified as mixed candidates.
Their values have been obtained using the p.d.f. from the simulation.

The functions ${\cal P}_{B_q}^{mix.}(t)$ and ${\cal P}_{B_q}^{unmix.}(t)$ have
been given in the introduction for neutral B-mesons. For charged B-mesons and $b$-baryons,
 the decay time distribution has a 
simple exponential behaviour. These distributions have to be convoluted
with the time resolution distributions ${\cal R}_{B}(t'-t)$ and 
${\cal R}_{BC}(t'-t)$ for direct and cascade semileptonic B decays 
respectively, obtained from the simulation.

For   $\Zz \rightarrow c \overline{c}$ and $\Zz \rightarrow$~light~quark pair
events, the reconstructed time distributions obtained in simulation have 
been fitted directly to provide  ${\cal P}_{C}(t)$ and ${\cal P}_{H}(t)$.

\begin{figure}[htb]
\begin{center}
\epsfig{figure=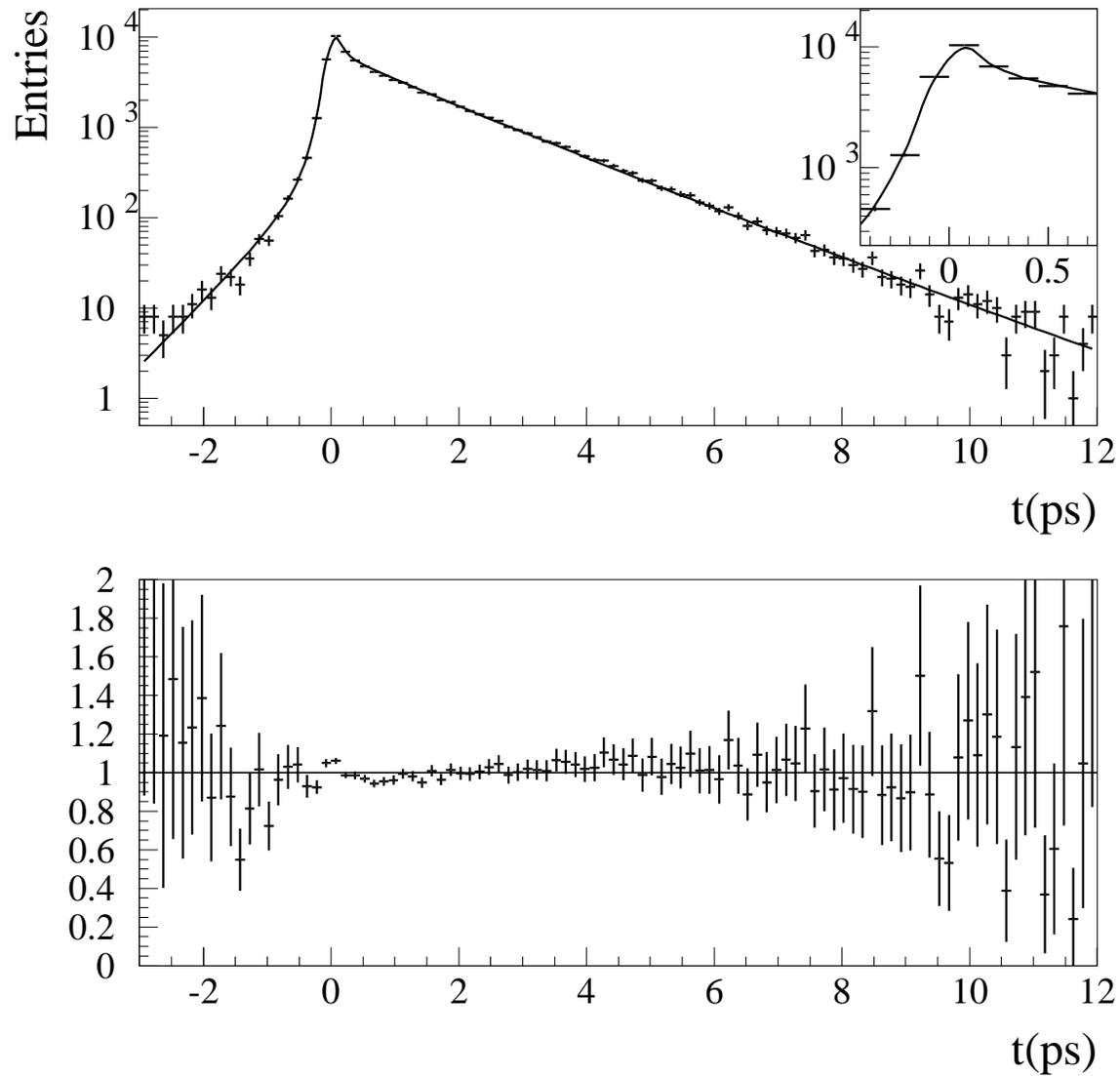,height=16cm}
\caption{  \it {  The plot on the top shows the proper-time distribution in real data (1994-2000) 
with the fit described in the text superimposed. 
The bottom plot is obtained dividing, bin-by-bin, the distribution of the 
proper time measured in data by the value of the fitted function. }}
\label{fig:resolincl_94}
\end{center}
\end{figure}

\begin{figure}[htb]
\begin{center}
\epsfig{figure=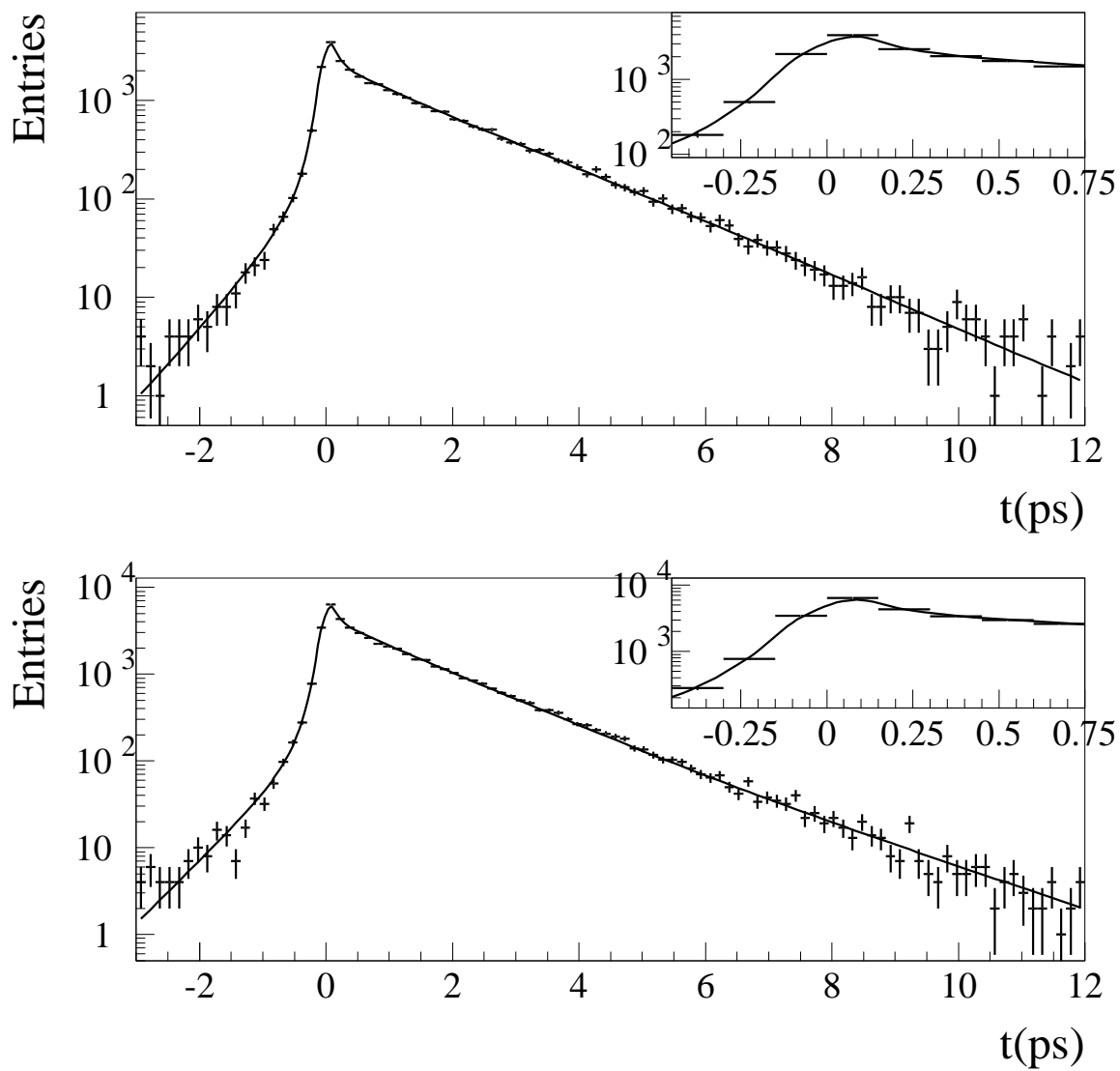,height=16cm}
\caption{  \it {  The plots on the top (bottom) show the proper-time distributions in real data (1994-2000) 
with the fit superimposed for like-sign (unlike-sign) events. } }
\label{fig:resolincl1_94}
\end{center}
\end{figure}

The proper decay time distribution of real data is shown
in Figure~\ref{fig:resolincl_94} for the 1994-2000 sample
with the result of the fit superimposed. 
The agreement obtained is compatible with the expected systematic 
uncertainty on the proper-time resolution.
The same distribution for like and unlike-sign events separately is shown in 
Figure~\ref{fig:resolincl1_94}.

The most relevant external inputs in the likelihood function are listed in
Table~\ref{tab:val}.

\section{  Results of the inclusive lepton analysis }
\label{res}
In this section the result of the measurement of the $B^0_s$ oscillation 
frequency, or a limit on its possible range, is given. 
The measurement of the $B^0_d$ oscillation frequency is also
presented as a cross-check of the analysis technique.

\subsection{Measurement of \boldmath $\Delta m_d$ }
\label{sec:261}

 An unbinned maximum likelihood method has been applied to the set of classified events.
The parameters corresponding to $\Delta m_d$ and to the p.d.f. for the tagging in the opposite hemisphere in four 
different intervals of the $x^{opp.}_{tag}$ discriminating variable have been fitted by minimizing the following function:

\begin{equation}
 {\cal L}~=~-\sum_{like-sign~ events} \ln({\cal P}^{like}(t))~-\sum_{unlike-sign~ events} \ln({\cal P}^{unlike}(t)).
\end{equation}

\noindent
The results, converted to tagging purities, 
are given in Table~\ref{tab:tabdata}.
The final result is:
$$
\Delta m_d~=~0.456 \pm 0.021 ~\mbox{ps}^{-1}.
$$
\begin{center}
\begin{table}[ht]
\begin{center}
\begin{tabular}{|c|c|c|c|c|}\hline
variable        &      1992-1993              &    1994-2000       &     $x^{opp.}_{tag}$ range     \\      \hline\hline
 $\dmd$~(ps$^{-1}$)    &  $0.459 \pm 0.036$     & $0.455 \pm 0.026$   &                    \\ \hline
 $\epsilon_b^1$ &  $0.550 \pm 0.008$          & $0.550 \pm 0.006$   &   [0.00-0.25]      \\ \hline
 $\epsilon_b^2$ &  $0.644 \pm 0.008$          & $0.661 \pm 0.006$   &   [0.25-0.50]      \\ \hline
 $\epsilon_b^3$ &  $0.786 \pm 0.008$          & $0.762 \pm 0.007$   &   [0.50-0.75]      \\ \hline
 $\epsilon_b^4$ &  $0.873 \pm 0.013$          & $0.889 \pm 0.008$   &   [0.75-1.00]      \\   \hline \hline
\end{tabular}
\caption{ \it {Results from the fit on real data of $\dmd$ and of the tagging purity in four intervals of the discriminating variable 
as indicated in the last column. The fit was performed assuming that $\Delta m_s$ is large ($\Delta m_s=20~\mbox{\rm ps}^{-1}$ has 
been used). }}
\label{tab:tabdata}
\end{center}
\end{table}
\end{center}

\begin{figure}[htb]
\begin{center}
\mbox{\epsfxsize=17cm \epsffile{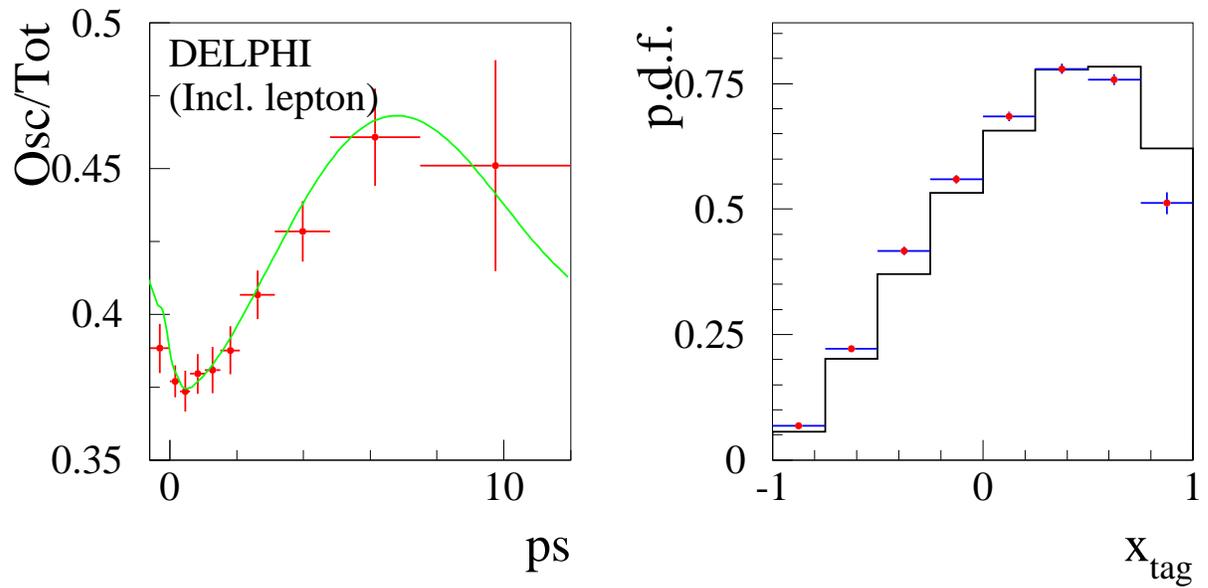}}
\caption{\it { Left plot: time dependence  of the fraction of like-sign events. The curve is the result of the log-likelihood fit. 
The right histogram shows the probability density function (p.d.f.) for tagging a $b$ quark from simulated
data. The points with the error bars show the result of the fit to real data.}}
\label{fig:dmdall}
\end{center}
\end{figure}

The experimental distribution of the fraction of like-sign events versus the 
decay time is shown in Figure~\ref{fig:dmdall}.
In the same figure the comparison between the fitted and simulated tagging 
p.d.f. is shown. 
The difference between these two p.d.f. shows that it is very
important, where possible, to compute 
the distributions of interest on the data.

The final tagging variable, $x_{tag}$, used in the $\Delta m_s$ analysis is 
obtained by combining the tagging variable
of the opposite hemisphere whose p.d.f. has been fitted
on the data (fixing the $\Delta m_d$ value to the world average) and the 
tagging variable of the same hemisphere whose p.d.f. has been computed 
from simulated data.

\begin{figure}[htb]
\begin{center}
\epsfig{figure=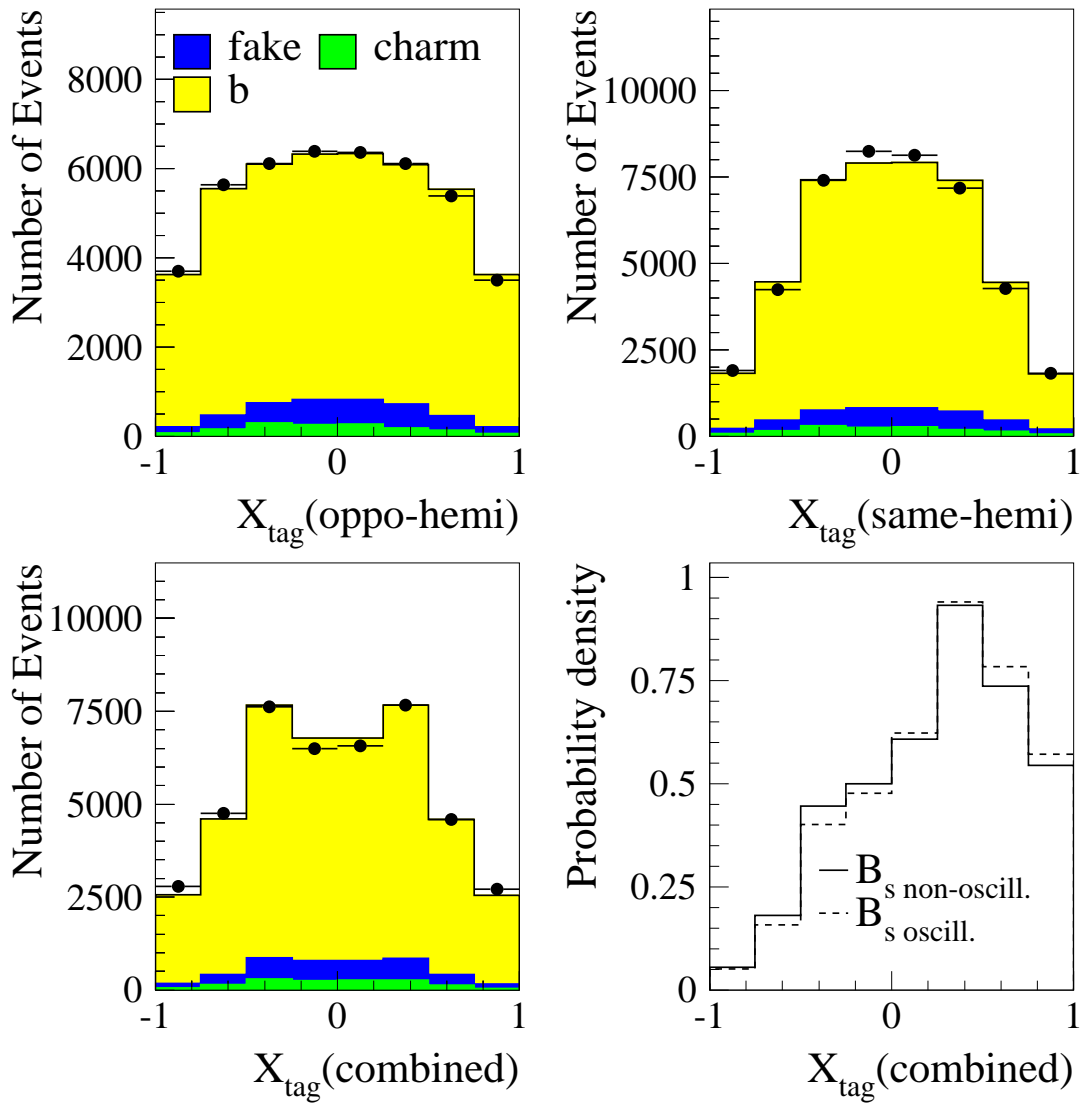,height=16cm}
\caption{\it { The plots show the overall comparison between the distributions 
of the tagging variables in data 
and the p.d.f. used in the fit for opposite (top-left), 
same (top-right) hemisphere and the combined tagging (bottom-left) 
for the $\mbox{B}^0_s - \overline{\mbox{B}^0_s}$ oscillation analysis. 
The plot on the bottom-right shows the expected p.d.f. for oscillating and 
non-oscillating $B^0_s$, separately. }}
        \label{fig:tagging}
\end{center}
\end{figure}

The overall agreement between the tagging p.d.f. used in the fit and the  
distributions of the tagging variable in the data is shown 
in Figure~\ref{fig:tagging}, together with 
the tagging p.d.f. for oscillating and non-oscillating $\Bs$
computed from simulated data.
The distributions for oscillating and non-oscillating
$\Bs$ are expected to coincide only if the tagging in the same hemisphere
is completely independent from the $\Bs$ oscillation; the small difference found in the
simulation has been taken into account in the fit by using different p.d.f.
for the oscillation and non-oscillation hypothesis.

\subsection{Limit on \boldmath $\dms$}
\label{sec:262}
The limit on $\dms$ has been obtained in the framework of the amplitude method.
In this method~\cite{ref:amplitude}, an oscillation amplitude, ${\cal A}$, 
is fitted for each assumed value of $\dms$.
The equations for ${\cal P}_{B_s}^{mix.}$ and ${\cal P}_{B_s}^{unmix.}$ 
become:
\bc
${\cal P}_{B_s^0}^{unmix.}~=~{\cal P}(\mbox{B}_{s}^{0}\rightarrow \mbox{B}_{s}^{0})~
=~\frac{1}{2 \tau_{s}} e^{- \frac{t}{\tau_{s}}}
 ~[ 1 + {\cal A}  \cos ({\Delta m_s t} ) ]$
\ec
and
\bc
${\cal P}_{B_s^0}^{mix.}~={\cal P}(\mbox{B}_{s}^{0}\rightarrow \overline{\mbox{B}_{s}^{0}})~
=~\frac{1}{2 \tau_{s}} e^{- \frac{t}{\tau_{s}}}
 ~[ 1 - {\cal A } \cos ({\Delta m_s t} ) ].$
\ec
For ${\cal A}=1$, the standard time distribution expressions for  
mixed and unmixed candidates given in section~\ref{sec:1} are recovered.
A measurement of the amplitude is obtained for each value of $\dms$.
It has been verified that the log-likelihood distribution 
has a parabolic behaviour around its minimum.

In this approach, it is also easy to compute the probability of excluding 
a given value of $\dms$ with the studied channel. 
It has to be assumed that the real value of $\dms$ is very large and in practice
not accessible with the present experimental sensitivity. 
The expected value of the amplitude is then equal to zero.
All measured values of ${\cal A}$ which satisfy 
${\cal A}~<~~1~-1.645~ \sigma_{\cal A}$ are such that the corresponding value 
of $\dms$ is excluded at the 95$\%$ C.L..
The $\dms$ value satisfying $1.645~ \sigma_{\cal A}=1$ 
is then defined as the sensitivity of the analysis. 

It is also possible, in this framework, to deduce the log-likelihood function $\Delta \log{\cal L}^{\infty}(\dms)$ 
referenced to its value obtained for $\dms=\infty$. The log-likelihood values can be easily deduced
from $\cal A$ and $\sigma_{\cal A}$ using the expressions given in~\cite{ref:amplitude}:
\renewcommand\arraystretch{1.1}
\begin{eqnarray}
\Delta \log{\cal L}^{\infty}(\dms) &
= \frac{\textstyle 1}{\textstyle 2}\,\left[\left({{\cal A}-
1 \over \sigma_{\cal A}}\right)^2-
 \left({{\cal A} \over \sigma_{\cal A}}\right)^2 \right]
& = \left (\frac{\textstyle 1}{\textstyle 2}
-{\cal A}\right){\textstyle 1\over 
   \textstyle \sigma_{\cal A}^2} 
\ , 
\label{dms_right1}\\
&\Delta \log{\cal L}^{\infty}({\dms})_{mix}   &= -\frac{1}{2}\frac{1}{\sigma_{\cal A}^2} 
\label{dms_right2} \ ,  \\
&\Delta \log{\cal L}^{\infty}({\dms})_{nomix} &=  \frac{1}{2}\frac{1}{\sigma_{\cal A}^2 }
\ . \label{dms_right3}
\end{eqnarray} 
\renewcommand\arraystretch{1.0}
The last two equations give the average log-likelihood value when
$\dms=\dms^{true}$ ($mixing$ case) and when $\dms$ is different from $\dms^{true}$ ($no$-$mixing$ case).\\

\begin{figure}[htb] 
\begin{center}
\epsfig{figure=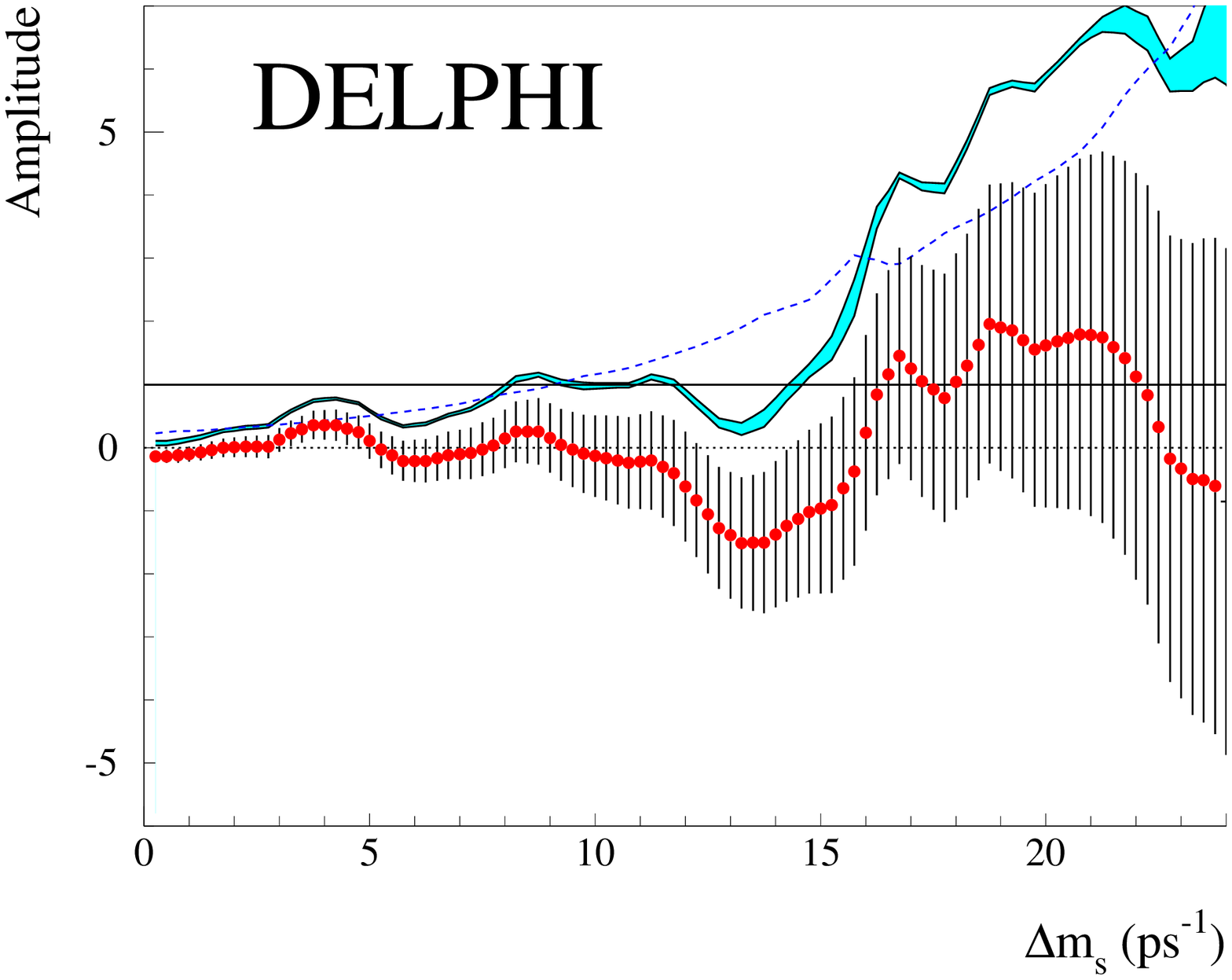,height=10cm}
\epsfig{figure=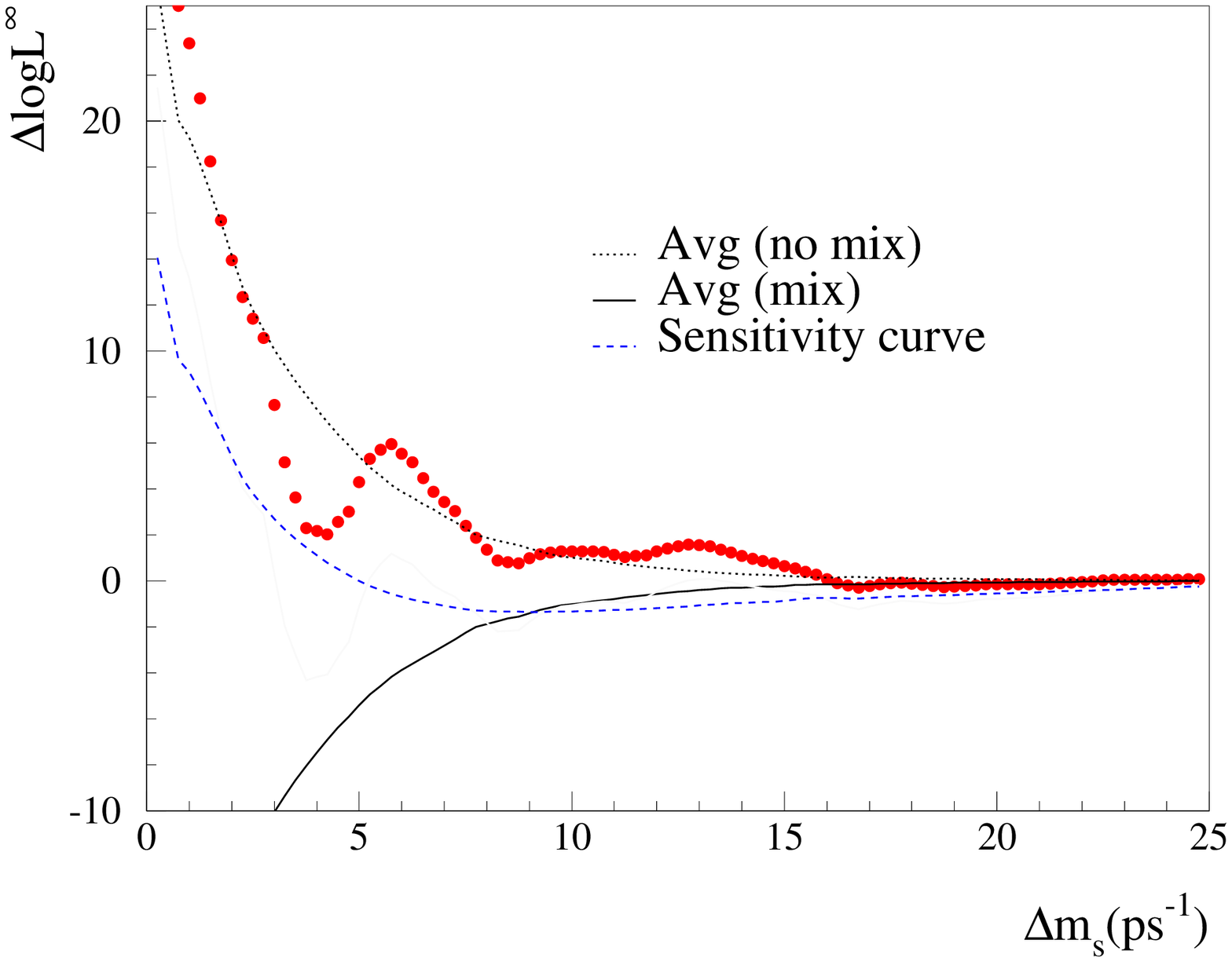,height=10cm}
\caption{ \it { 
DELPHI inclusive lepton analysis: 
the upper plot shows the variation of the oscillation  amplitude ${\cal A}$ as a 
function of $\Delta m_{s}$. 
The filled area shows the variation of the contour corresponding to 
${\cal A} + 1.645 \sigma_{\cal A}$ when the systematic uncertainty is included.
The dotted line shows the sensitivity. 
The likelihood referenced to $\dms=\infty$ (lower plot), represented by points, has 
been deduced from the amplitude spectrum using the formula given 
in~\cite{ref:amplitude} (see section~\ref{sec:262}). }}
\label{fig:dmsincl_all}
\end{center}
\end{figure}

Using the amplitude approach (Figure~\ref{fig:dmsincl_all}), and considering only the statistical uncertainties, a 95$\%$ C.L limit
on $\Delta m_s$ was set:
\begin{eqnarray}
   \begin{array}{ll}
\dms > 8.0 ~ \mbox{ps}^{-1}~\mbox{at the 95\% C.L.} & \\
\mbox{\rm {with a sensitivity of}}~~ \dms = 9.6 ~ \mbox{ps}^{-1}.
   \end{array}
 \end{eqnarray}

\begin{figure}[htb] 
\begin{center}
\begin{tabular}{cc}
\epsfig{figure=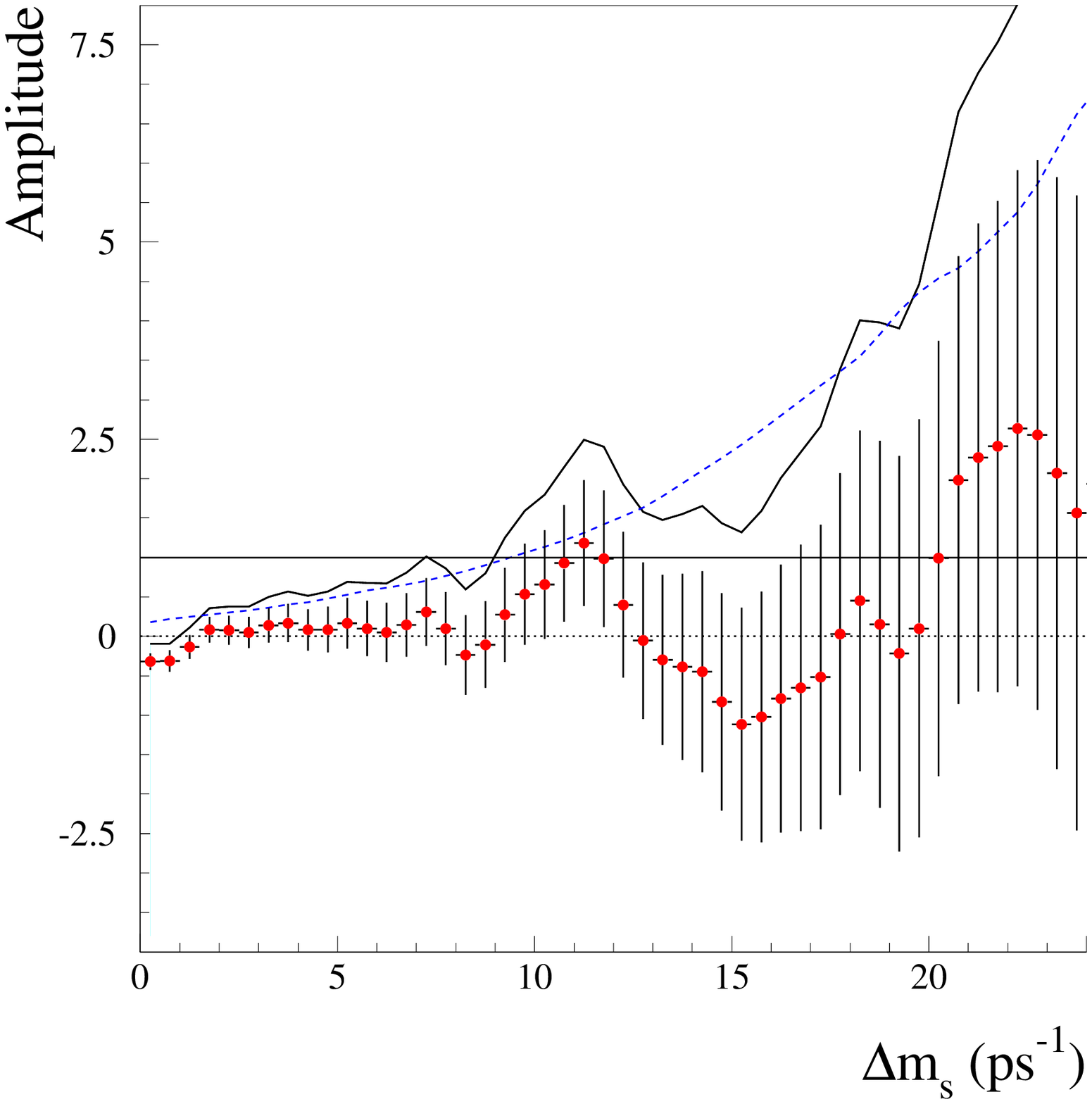,width=7.5cm} &
\epsfig{figure=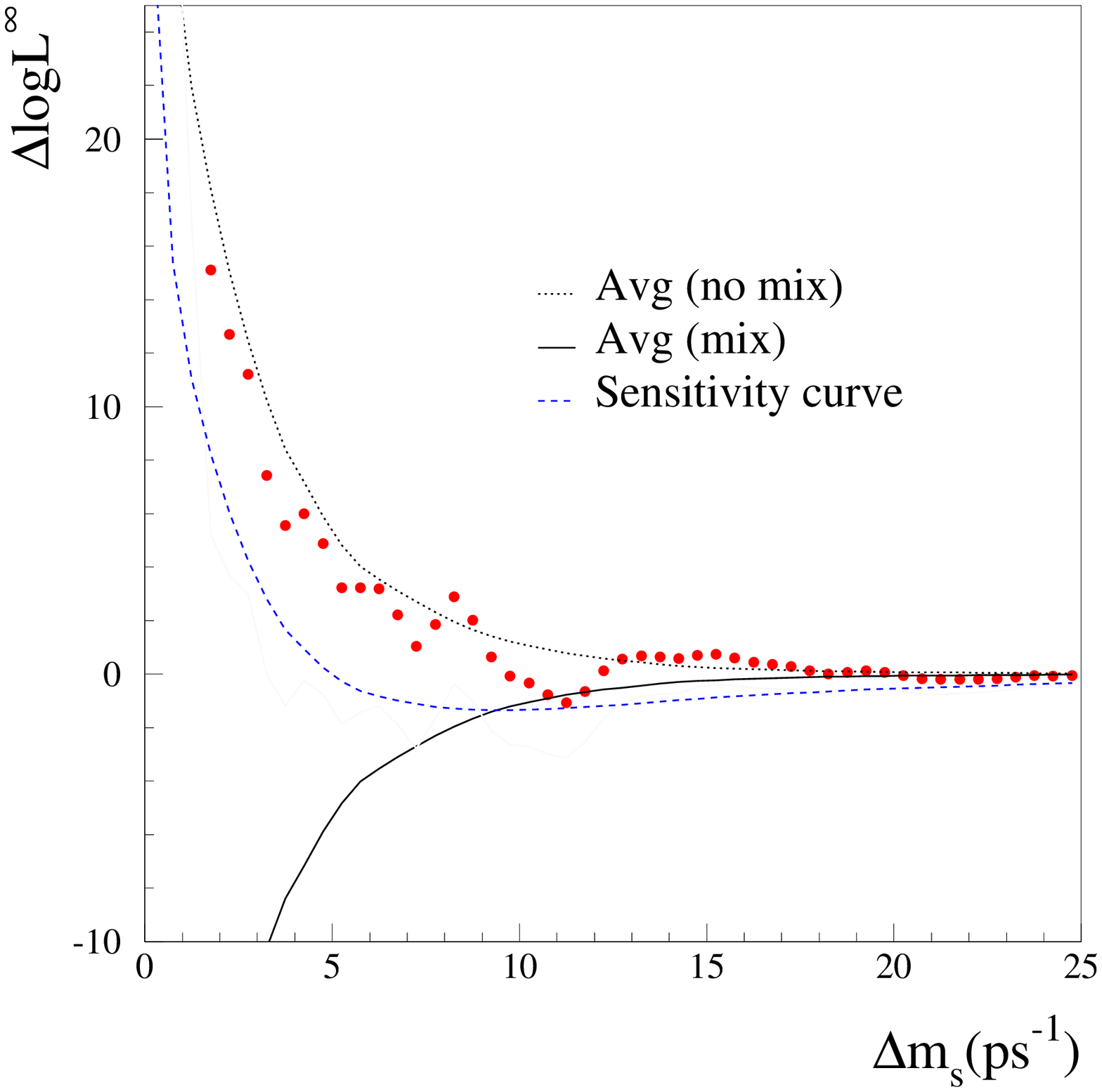,width=7.5cm} \\
\epsfig{figure=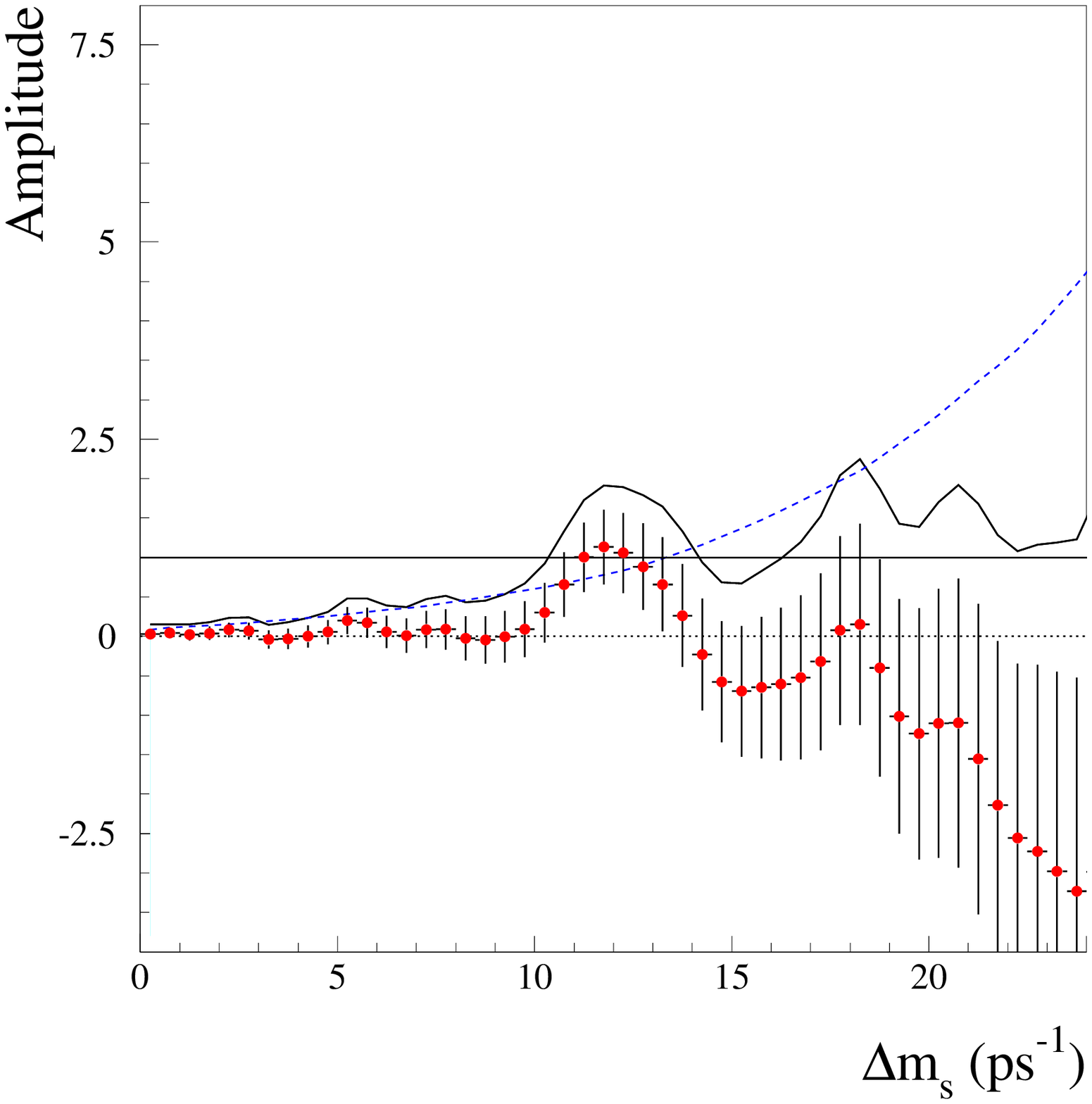,width=7.5cm} &
\epsfig{figure=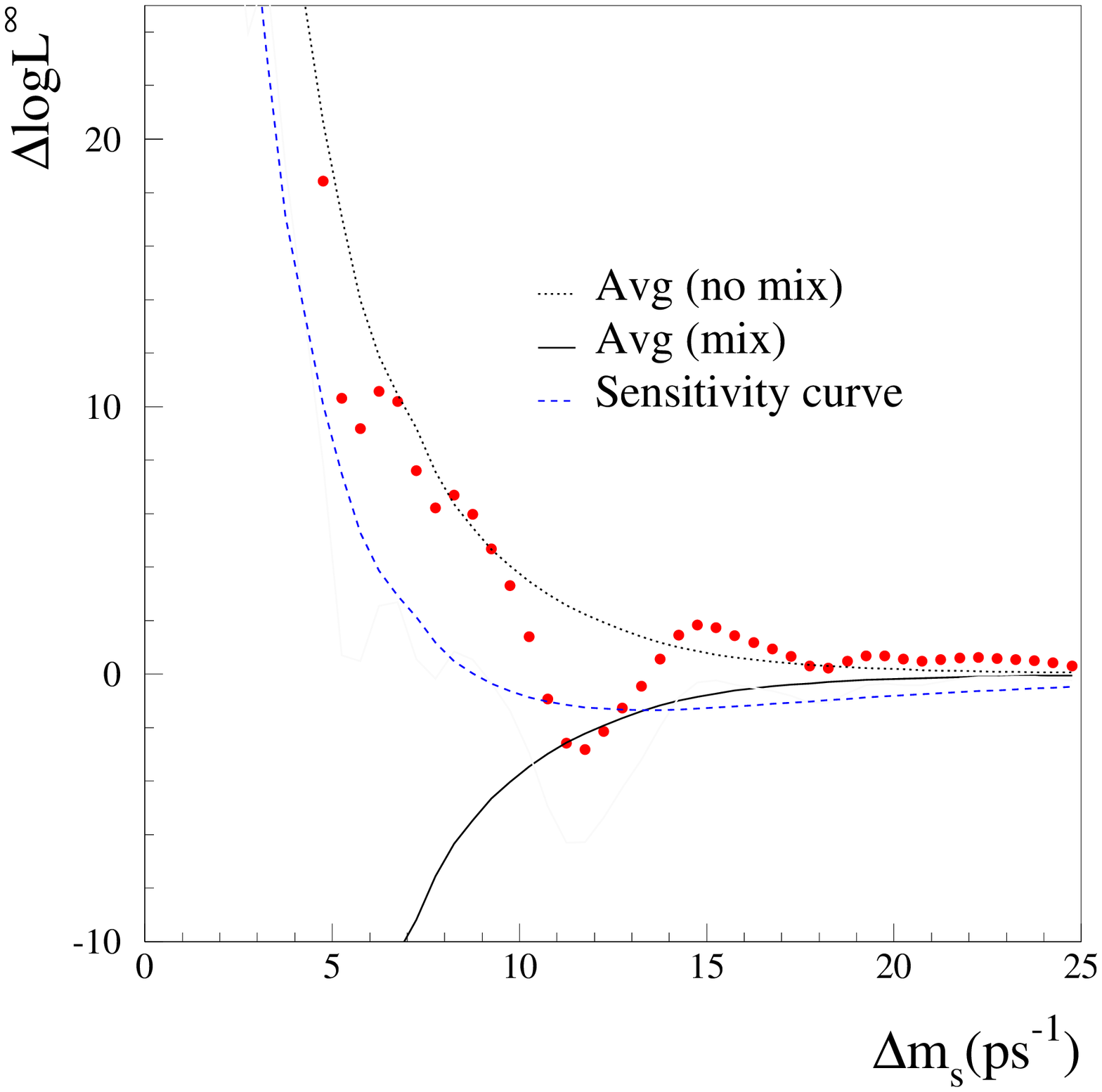,width=7.5cm}
\end{tabular}
\caption{ \it  Check of the overall fit procedure using simulated data. 
The plots on the top show the variation of the oscillation amplitude ${\cal A}$ 
as a function 
of $\Delta m_{s}$ (left) and the likelihood referenced to $\dms=\infty$ (right) 
for the simulated data with standard $\Bs$ purity. 
The plots on the bottom give the same information if the $\Bs$ purity is 
increased to 30$\%$. 
The signal is expected (and seen) at 11.2 ps$^{-1}$.}
\label{fig:chebello}
\end{center}
\end{figure}

A check of the overall procedure has been performed using simulated data 
(Figure~\ref{fig:chebello}) 
with different values of the $\Bs$ purity.
It can be seen that the minimum of the likelihood 
corresponds to the generated value of $\dms$ 
and that its significance increases with the $\Bs$ purity.
This check gives some confidence that the Monte Carlo parameterisations 
are correct.

\subsubsection{Systematic uncertainties on \boldmath $\dms$} 
\label{p:dms_syst}
Systematics have been evaluated by varying, according to their
respective uncertainties, the values of the parameters which were kept constant
in the evaluation of the log-likelihood function. For each parameter, the
variation of the amplitude and of its measurement error were taken into account
in the evaluation of the systematic uncertainty. This was done in the following 
way~\cite{ref:amplitude}:
\begin{eqnarray}
  \label{eqrsysa}
  \sigma_{\cal A}^{sys} & = & {\cal A - A}_i + (1-{\cal A}_i)\frac{\sigma_{\cal
  A}^{stat}-\sigma_{{\cal A}_i}^{stat}}{\sigma_{{\cal A}_i}^{stat}}
\end{eqnarray}
where ${\cal A}$, $\sigma_{\cal A}$ (${\cal A}_i$, $\sigma_{{\cal A}_i}$) 
indicate the values 
and the errors on the amplitude after (before) the parameter variation; 
$\sigma^{stat}$ is the statistical error on the amplitude. 
The following systematics have been considered:

\begin{itemize}
\item {\it  $\dmd$, $\epsilon^{i}_b$ }. \\
The systematic error due to $\dmd$ has been evaluated by varying the
central value given in Table~\ref{tab:val} by its standard deviation.

The tagging purity in the opposite hemisphere, $\epsilon^{i}_b$,
has been obtained from a fit to the real data 
(see section~\ref{sec:261} and Table~\ref{tab:tabdata}). 
The systematics have been evaluated by varying the shape of the variable of
the opposite hemisphere tagging in accordance with the fitted errors 
in a way to maximize the data and Monte Carlo agreement 
in Figure~\ref{fig:tagging}.
The shape has been modified using a single parameter whose effect, on the
mean purity, is $\pm 0.005$ ($\pm 0.007$) in the 1992-1993 (1994-2000) data.

A conservative variation of the shape of the same hemisphere tagging 
corresponding to an integral variation of $\pm 0.010$ has been also
added.


\item {\it Fractions of leptons. } \\
A relative variation of $\pm 20\%$ has been applied on the fraction of fake 
leptons, leptons from charm and cascade semileptonic decays. 
This variation is equal to the uncertainty on
the measurement of the fake lepton rate in data. 
These changes in the parameters have been compensated by a corresponding 
variation of the fraction of direct leptons.

The systematics coming from the uncertainties on the shape of the 
discriminating variable distribution have been evaluated by forcing, 
with appropriate reweighting, an exact
match between the distributions in data and simulation 
in Figure~\ref{fig:discrbl_94}.

\item {\it Production rates of B hadrons. } \\
Following the procedure described in section~\ref{sec:24} the value of 
$f_{B_s}$ was varied inside its measured error. \\
The systematics coming from the uncertainties on the shape of the 
discriminating variable distribution has been evaluated by forcing, 
with appropriate reweighting, an exact
match between the distributions in data and simulation
in Figure~\ref{fig:discrbs_94}.

\item {\it Systematics from the resolution of the B decay proper time. } \\
If the errors are Gaussian, the oscillation amplitude is damped by a factor 
$\rho$ because of the finite  accuracy in the decay time $ \sigma_t$:
\begin{equation}
\rho~=~ e^{-(\dms \sigma_t)^2/2}
\end{equation}
where $ \sigma_t$ receives two contributions: 
from decay distance and from momentum measurements.

At small decay times, the accuracy on $t$ depends mainly on the error on 
the decay distance. This quantity was measured 
using simulated events, after tuning the track reconstruction 
efficiencies and measurement errors to match the real data.
For this purpose, charged particles emitted at angles smaller than 30$^0$ 
from the horizontal plane have been selected, in order to benefit from
the precise definition of the beam position in the vertical direction. 
The details of the tuning procedure are described in~\cite{ref:borissov}. 
After the tuning, the agreement between real and simulated data on the decay 
distance error has been 
evaluated from the width of the negative part of the flight distance 
distribution, for events which are depleted in b-hadrons.
The difference of the widths in  real data and in simulation is found to be 
approximately 10\%. 

The uncertainty on the momentum resolution was estimated to be $\pm$10\%.
This number was obtained by  comparing the reconstructed B momentum
in a hemisphere with the expected momentum in that hemisphere,
obtained using energy and momentum conservation, for data and simulation;
it was found that the momentum resolution agreed to better than $\pm$10\%.

The systematic error coming from the uncertainties on the resolution functions
is evaluated by varying by $\pm10\%$ the two parameters describing the linear time
dependence of the narrower Gaussian (see Table~\ref{tab:tabpart}).
A variation of $\pm10\%$ of the resolution for the background events is also 
considered. 

An important check has been carried out to verify that the subdivision in 
resolution classes 
does not introduce differences between the real and simulated data:
the variation of the amplitude error as a function 
of $\Delta m_{s}$ for real and simulated data has been compared. 
After rescaling data and Monte Carlo simulation to the same number 
of $\Bs$ events, the ratio of the amplitude errors is compatible
with 1 within  $\pm10\%$. An extra systematic uncertainty has been added by allowing a $\pm10\%$ variation 
of the amplitude error.
\end{itemize}

The previous quantities have been varied separately. Table~\ref{tab:syst} summarizes the various contributions to the systematic error for different $\Delta m_{s}$ values. The global systematic error is obtained by taking the quadratic sum of 
the individual systematics.

\begin{center}
\begin{table}[ht]
\begin{center}
\begin{tabular}{|l|c|c|c|c|}\hline
 
 \mbox{Contribution to the systematic error}  & $5~ps^{-1}$
                     & $10~ps^{-1}$
                     & $15~ps^{-1}$
                     & $20~ps^{-1}$ \\      \hline\hline
  \mbox{$f_{\small \Bs}$}                & 0.04 & 0.03 & 0.03 & 0.02 \\ \hline
  \mbox{Fake lepton fraction}   & 0.02 & 0.02 & 0.02 & 0.02 \\ \hline
  \mbox{Lepton composition (p.d.f. shape)}   & 0.01 & 0.02 & 0.05 & 0.04 \\ \hline
  \mbox{B-hadron fractions (p.d.f. shape)}     & 0.01 & 0.02 & 0.03 & 0.03 \\ \hline
  \mbox{Tagging purity    (p.d.f. shape)}    & 0.01 & 0.01 & 0.02 & 0.03 \\ \hline
  \mbox{Flight length resolution}            & 0.02 & 0.07 & 0.31 & 0.33 \\ \hline
  \mbox{Momentum resolution}                 & 0.01 & 0.04 & 0.12 & 0.13 \\ \hline
  \mbox{Background proper-time shape}        & 0.01 & 0.01 & 0.01 & 0.01 \\ \hline
  \mbox{Subdivision in resolution classes}    & 0.03 & 0.07 & 0.14 & 0.26 \\ \hline\hline
  \mbox{Total systematic error}              & 0.14 & 0.28 & 0.68 & 0.72 \\ \hline\hline
  \mbox{Statistical error}                   & 0.30 & 0.65 & 1.35 & 2.56  \\ \hline
\end{tabular}
\caption{ \it {List of the various contributions to the systematic error for different $\Delta m_{s}$ values.}}
\label{tab:syst}
\end{center}
\end{table}
\end{center}

Including the present 
evaluation of systematics in the measured amplitude, the 95$\%$ C.L. limit is:
\begin{eqnarray}
   \begin{array}{ll}
~~~~~~~~~~\dms > 8.0 ~ \mbox{ps}^{-1}~\mbox{at the 95\% C.L.} & \\
~~~~~~~~~\mbox{\rm {with a sensitivity of}}~~ \dms = 9.1 ~ \mbox{ps}^{-1}.
   \end{array}
 \end{eqnarray}

\section {Update of the \boldmath $\mathrm{D_s^{\pm}}\ell^\mp$ analysis }

\begin{figure}[htb] 
\begin{center}
\begin{tabular}{cc}
\epsfig{figure=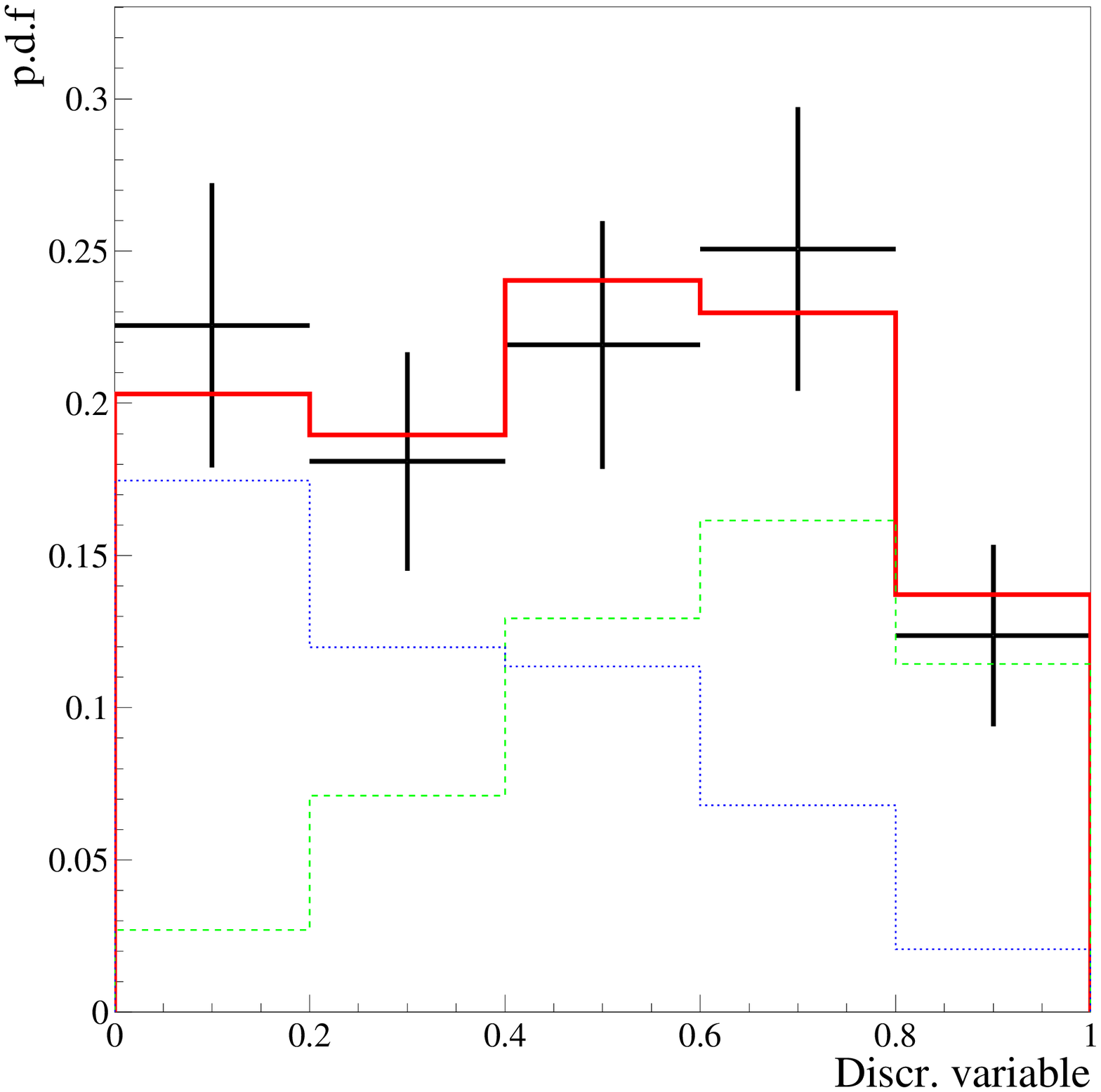,height=7.5cm} &
\epsfig{figure=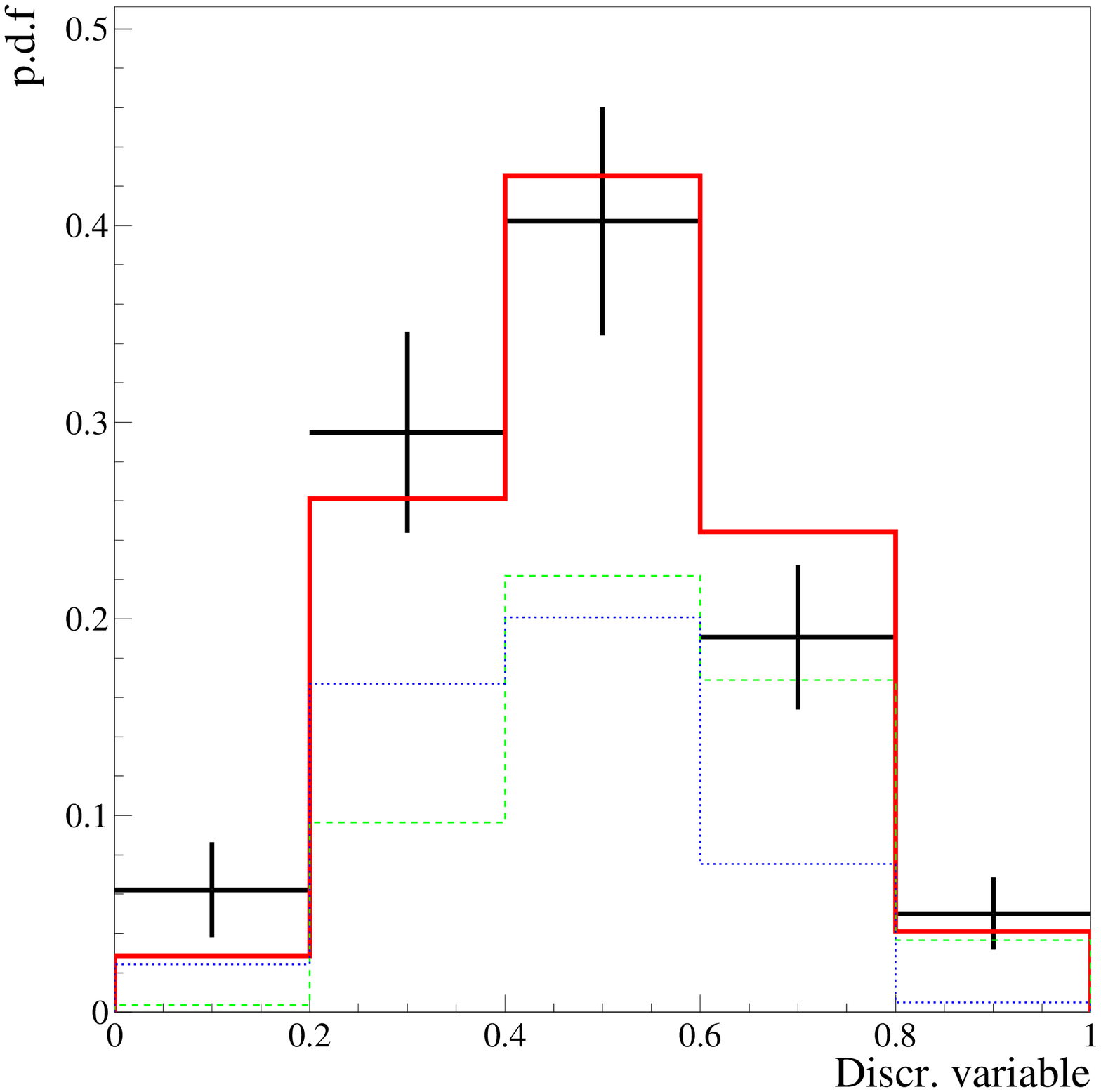,height=7.5cm} \\
\epsfig{figure=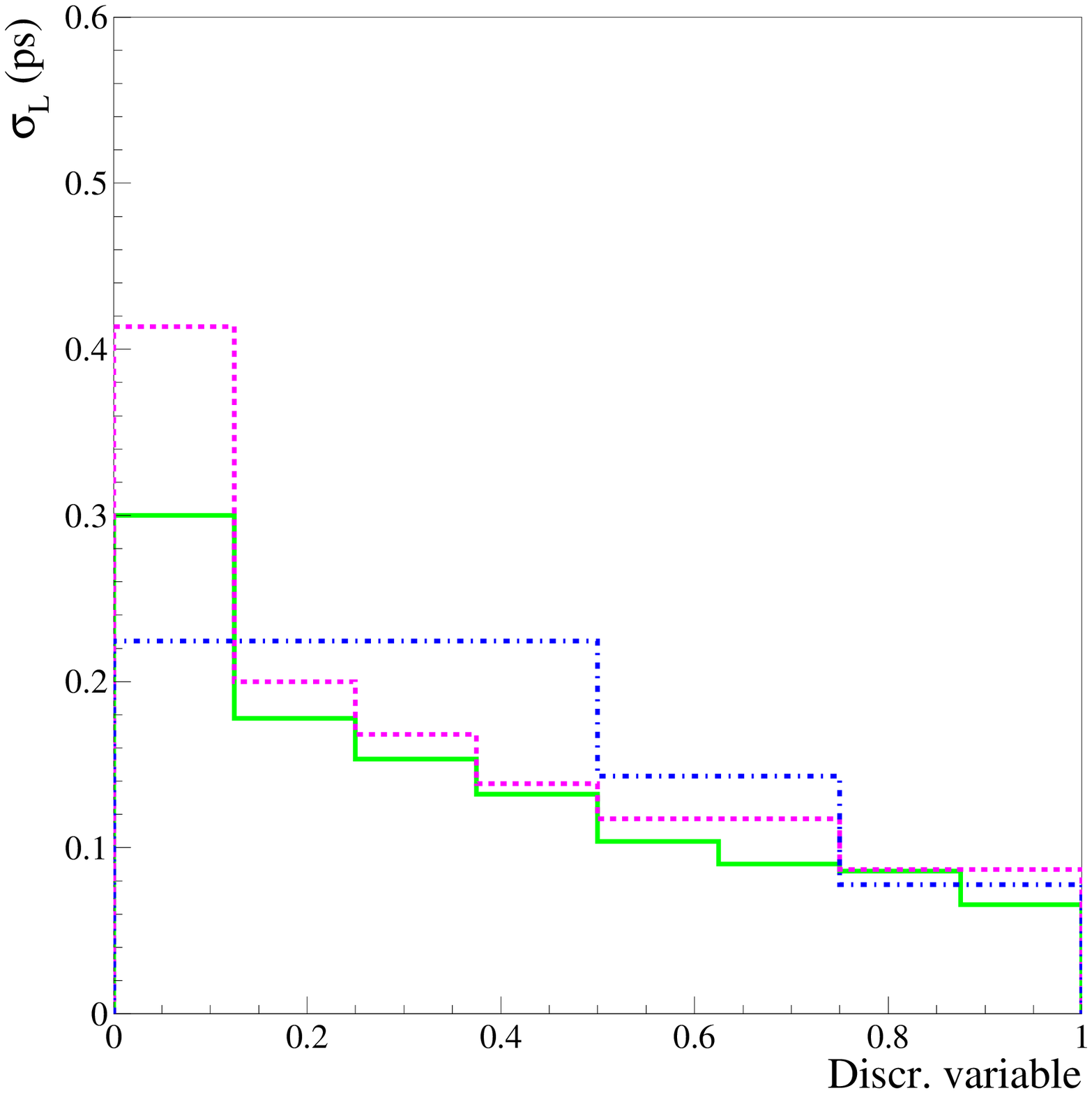,height=7.5cm} &
\epsfig{figure=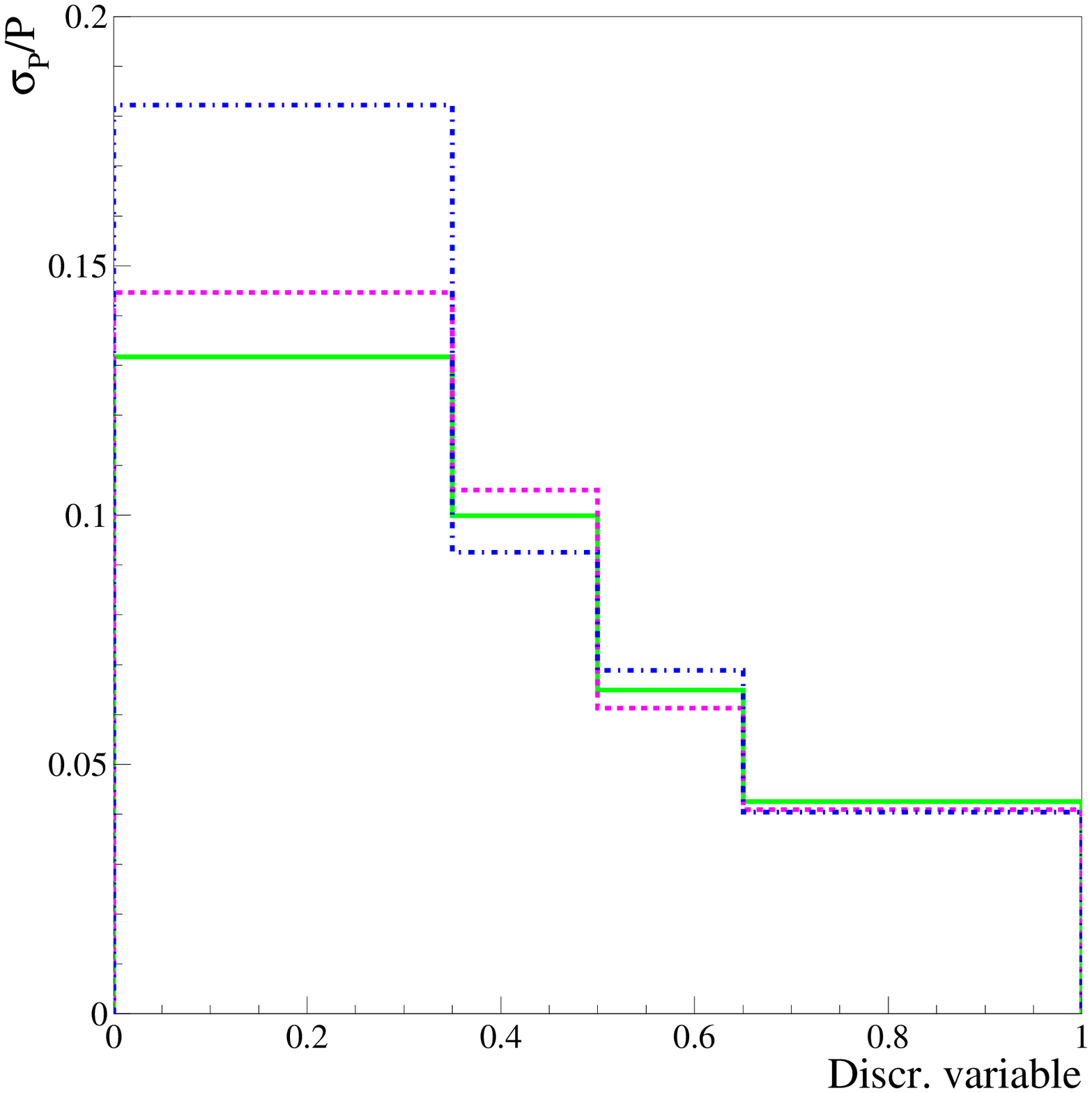,height=7.5cm}
\end{tabular}
\caption{ \it { Data/Monte Carlo comparison (represented with error bars
 and full line histogram respectively) for the discriminant variable 
for $\sigma_L$ (upper left) (the events having a resolution better (worse) 
than 250~$\mu$m (dashed histograms) have a value nearer to 1 (0)) 
and for $\sigma_p/p$ (upper right)
(the events having a resolution better (worse) than 8$\%$  (dashed histograms)
have a value nearer to 1 (0)). 
The lower plots show the evolution of $\sigma_L$ and $\sigma_p/p$ 
as a function of the discriminant variable. 
The continuous line is for the $\phi \pi$ and $K^{\ast}K$
decay modes, the dot-dashed line is for the $\phi \ell \nu_{\ell}$ decay mode and
the dashed line is for the $K^0K$ decay mode. Only the width of the narrowest Gaussian
is shown.
}}
\label{fig:sigmapara}
\end{center}
\end{figure}

 $\Bs-\Bsb$ oscillations have also been studied
using an exclusively reconstructed $\Ds$ meson correlated with
a lepton of opposite charge emitted in the same hemisphere:
\begin{eqnarray*}
\overline{\Bs} \longrightarrow {\mathrm {D_s^+}} \ell^- \overline{\nu_{\ell}} X.
\end{eqnarray*}
Details of the analysis are described in~\cite{ref:dslold}. A
limit $\mbox{at the 95\% C.L.}$ on the mass difference 
between the physical $\mbox{B}^0_{\mathrm s}$ 
states was obtained to be $\dms > 7.4 ~ \mbox{ps}^{-1}$, 
with a sensitivity of $\dms = 8.4 ~ \mbox{ps}^{-1}$. 
The weak point of this analysis was the rapid increase of
the error on the amplitude as a function of $\dms$. It was due to a global 
parameterisation of the proper-time resolution.
An improvement is obtained here by using the 
information of the proper-time resolution on an event-by-event basis.
A discriminant analysis is performed to separate B secondary vertices with an
expected good resolution (better than 250~$\mu$m). 
This discriminant variable is constructed by using:
\begin{itemize} 
\item  the error on the reconstructed decay length for the B and the D 
candidate;
\item  the $\chi^2$ of the B and the D secondary vertices;
\item  the D decay length divided by its error.
\end{itemize}
Another discriminant variable has been constructed to separate events having a 
momentum resolution better or worse than 8$\%$.
The variables used are:
\begin{itemize} 
\item  the reconstructed momentum;
\item  the $\Ds$ mass.
\end{itemize}
The two discriminant variables and their variation versus the 
error due to the decay length resolution, $\sigma_L=m_B/p_B\sigma(d_B)$, and  relative momentum error, 
$\sigma_{p_B}/p_B$, are shown in Figure~\ref{fig:sigmapara}. 

The analysis is repeated for all the decay channels and separately
for the 1992-1993 and 1994-1995 data.
The proper-time resolution,
$\sigma_t^2 = \sigma_L^2 + t^2 (\sigma_p/p)^2$, 
is then obtained for each pair
of values of the two discriminant variables and used in the fit procedure on an 
event-by-event basis. 

A check has been done to  
verify, using simulated data, the new treatment of the proper-time resolution.
The resolution obtained on an event-by-event basis by summing up all the
contributions from the Gaussian distribution of widths $\sigma_t$ 
reproduces well the overall resolution.

Another check has been performed by comparing the variation of the 
amplitude error, $\sigma_{\cal A}$, as a function of $\Delta m_{s}$ for real 
and simulated data. 
The agreement found indicates that the use of the proper-time 
resolution on an event-by-event basis does not introduce any significant 
difference in the error on the amplitude 
between real and simulated data. 
An extra systematic uncertainty is added by allowing a $\pm10\%$ variation of the amplitude 
error.

\begin{figure}[htb] 
\begin{center}
\epsfig{figure=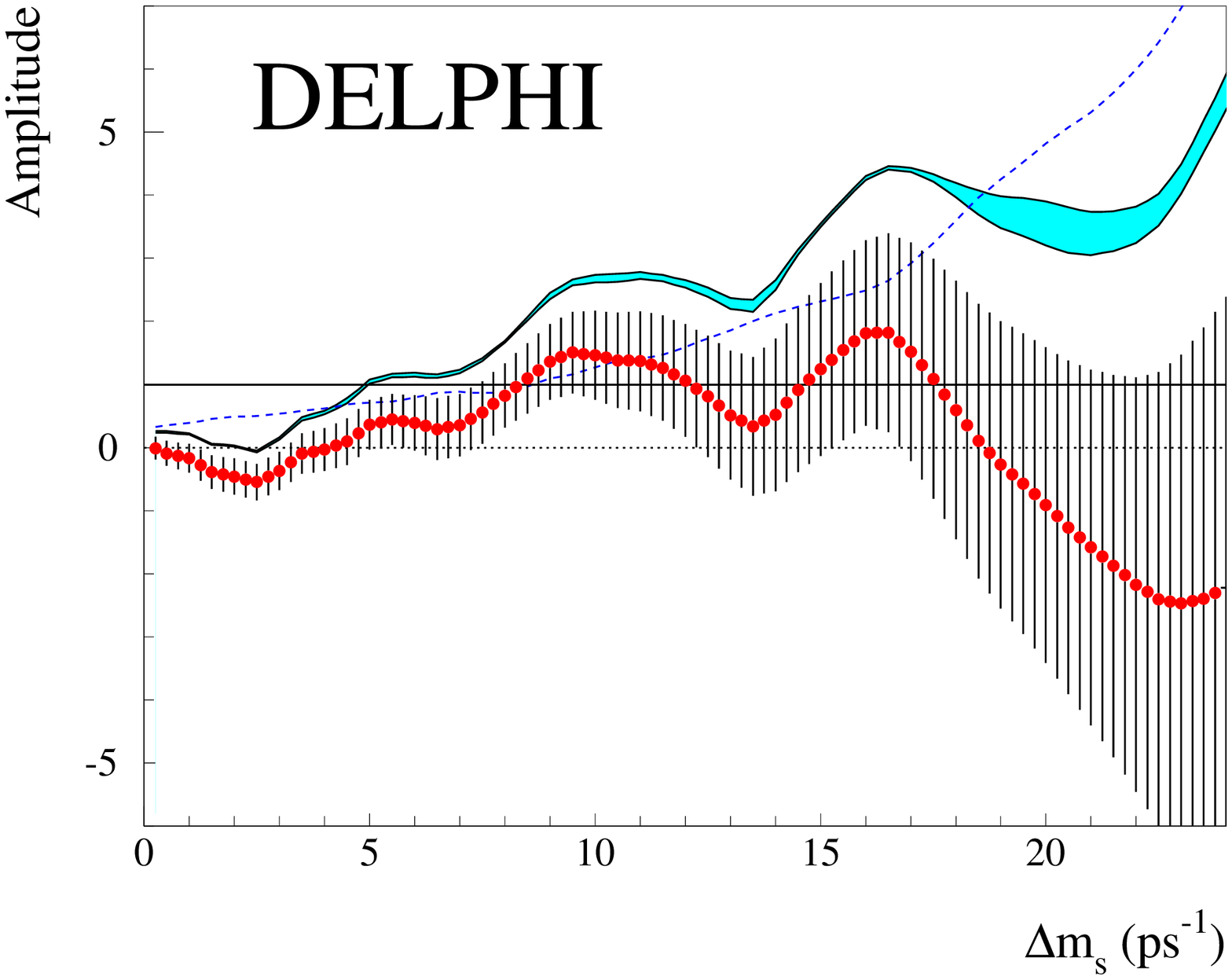,height=10cm}
\epsfig{figure=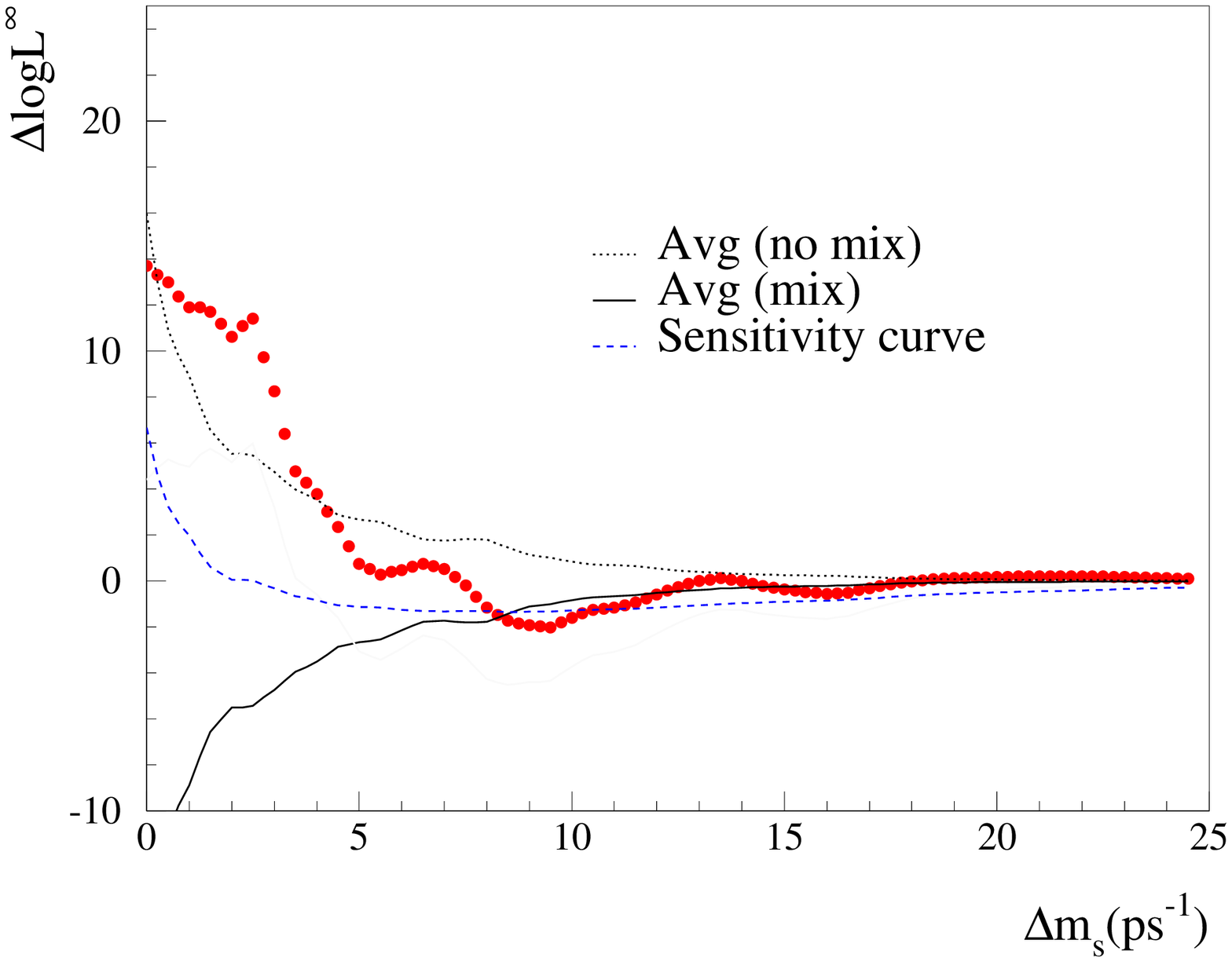,height=10cm}
\caption{ \it { 
DELPHI ${D_s^{\pm}}\ell^\mp$ analysis: 
the upper plot shows the variation of the oscillation  amplitude ${\cal A}$ as a 
function of $\Delta m_{s}$. 
The filled area shows the variation of the contour corresponding to 
${\cal A} + 1.645 \sigma_{\cal A}$ when the systematic uncertainty is included.
The dotted line shows the sensitivity. 
The likelihood referenced to $\dms=\infty$ (lower plot), represented by points, has 
been deduced from the amplitude spectrum using the formula given 
in~\cite{ref:amplitude} (see section~\ref{sec:262}). }}
\label{fig:dms_dsl}
\end{center}
\end{figure}

Using the amplitude approach (Figure~\ref{fig:dms_dsl}) a 95$\%$ C.L limit on 
$\Delta m_s$ was set:
\begin{eqnarray}
   \begin{array}{ll}
\dms > 4.9 ~ \mbox{ps}^{-1}~\mbox{at the 95\% C.L.} & \\
\mbox{\rm {with a sensitivity of}}~~ \dms = 8.6 ~ \mbox{ps}^{-1}.
   \end{array}
 \end{eqnarray}
The error on the amplitude at 20.0 $\mbox{ps}^{-1}$ has been decreased by a 
factor~1.7 from that presented in~\cite{ref:dslold} and also the sensitivity
has improved.

\section{Summary and combined limit on \boldmath $\dms$}
\label{sec:3}

Using data registered with the DELPHI detector between 1992 and 2000 and 
considering the
correlation between the sign of the charge of a lepton emitted at large 
transverse momentum relative to its jet axis, with the sign of a discriminating
variable which uses several parameters of the event to define
mixed and unmixed candidates, the value of the mass difference between B$^0_d$ 
mass eigenstates has been measured to be:
\begin{equation}
\Delta m_d~=~0.456 \pm 0.021 ~\mbox{ps}^{-1}
\end{equation}
where the error accounts only for the statistical uncertainty.
Using the amplitude approach, a 95$\%$ C.L. limit on $\Delta m_s$ was set:
\begin{eqnarray}
   \begin{array}{ll}
\dms > 8.0 ~ \mbox{ps}^{-1}~\mbox{at the 95\% C.L.} & \\
\mbox{\rm {with a sensitivity of}}~~ \dms = 9.1 ~ \mbox{ps}^{-1}.
   \end{array}
 \end{eqnarray}

The previously published $\dms$ analysis based on $\Ds$-lepton 
events~\cite{ref:dslold}
has been updated improving the treatment of the proper-time resolution
and the sensitivity.
Using the amplitude approach, a 95\% C.L. limit on $\Delta m_s$ has been set:
\begin{eqnarray}
   \begin{array}{ll}
\dms > 4.9 ~ \mbox{ps}^{-1}~\mbox{at the 95\% C.L.} & \\
\mbox{\rm {with a sensitivity of}}~~ \dms = 8.6 ~ \mbox{ps}^{-1}. 
   \end{array}
 \end{eqnarray}

\begin{figure}[htb] 
\begin{center}
\epsfig{figure=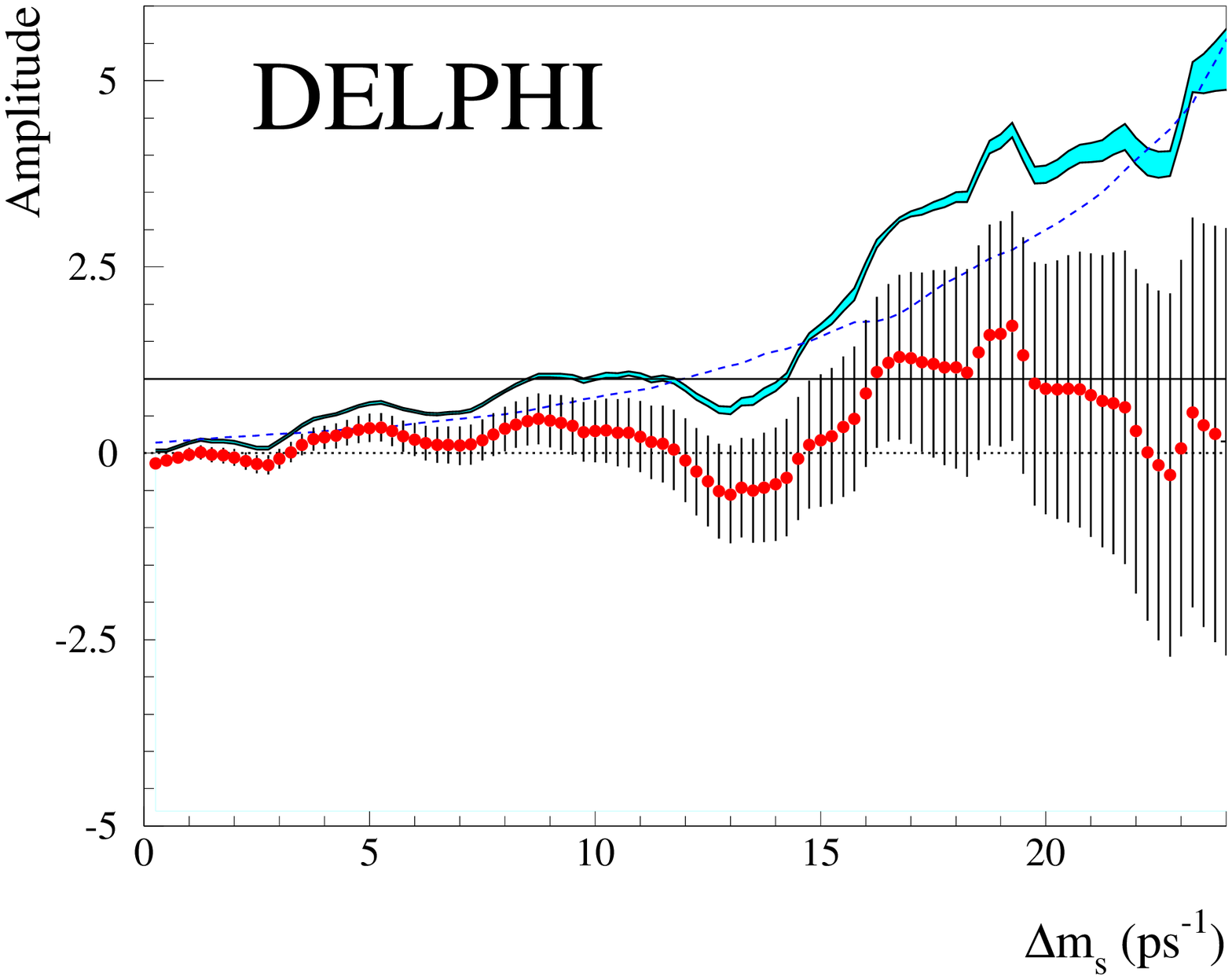,height=10cm}
\epsfig{figure=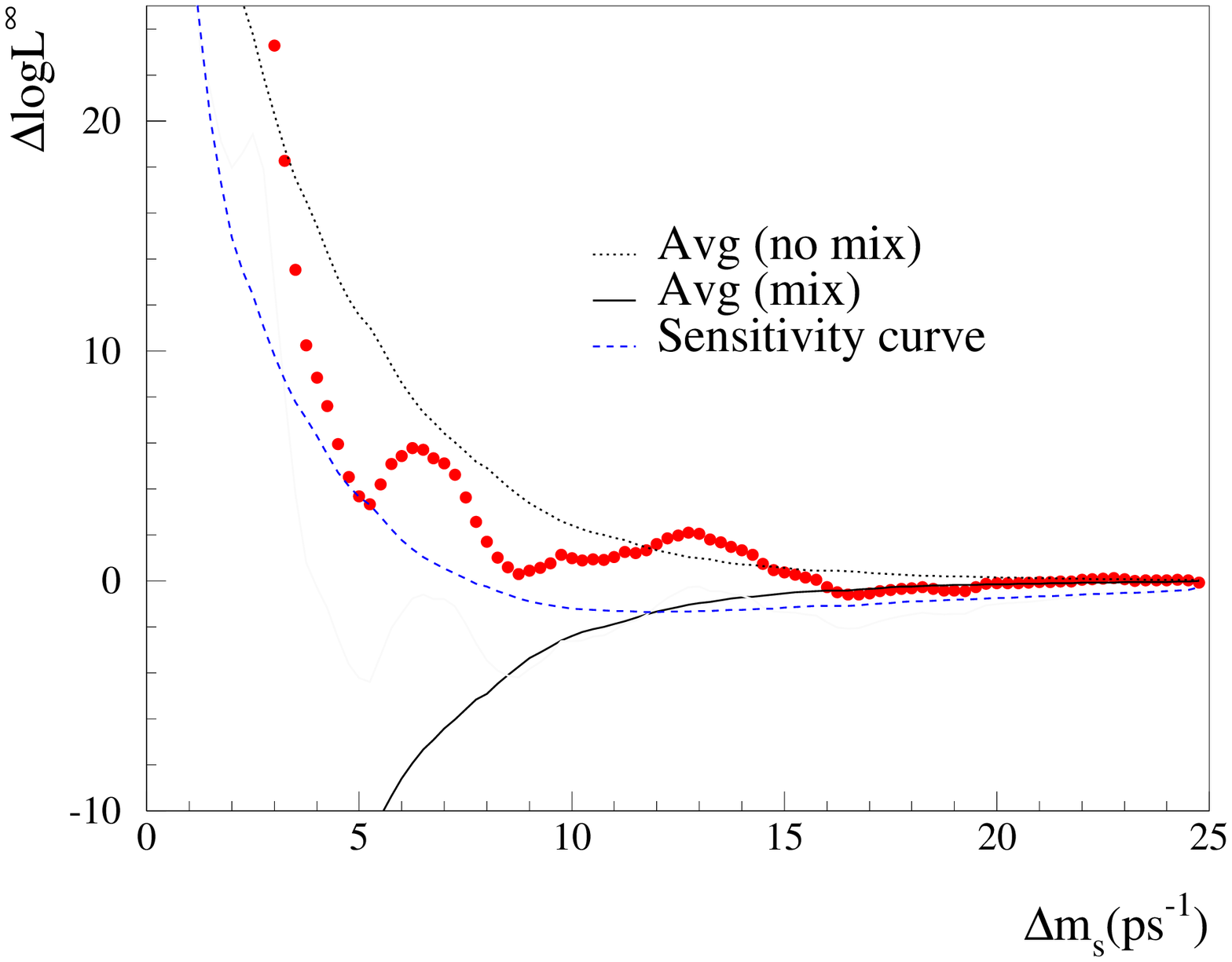,height=10cm}
\caption{ \it {DELPHI combined analysis: 
variation of the oscillation  amplitude ${\cal A}$ as a function of 
$\Delta m_{s}$. 
The filled area shows the variation of the contour corresponding to 
${\cal A} + 1.645 \sigma_{\cal A}$ when the systematic uncertainty is included.
The dotted line shows the sensitivity. 
The likelihood referenced to $\dms=\infty$ (lower plot), 
represented with points, has 
been deduced from the amplitude spectrum using the formula given 
in~\cite{ref:amplitude} (see section~\ref{sec:262}). }}
\label{fig:figcomb}
\end{center}
\end{figure}

DELPHI has performed three other $\dms$ analyses using 
exclusively reconstructed $\mbox{B}^0_s$ mesons~\cite{ref:exclbs}, 
$\Ds$-hadron events~\cite{ref:exclbs}, 
and the inclusively reconstructed vertices~\cite{ref:dipole}.
These analyses on $\Delta m_s$ have been combined, taking into account correlations 
between systematics in the different amplitude measurements 
(Figure~\ref{fig:figcomb}). The inclusive
vertices analysis, the inclusive lepton analysis and the $\Ds \ell$ analysis 
are statistically uncorrelated (high $p_T$ leptons have been
excluded from the vertex sample, $\Ds \ell$ events have been excluded
from the inclusive lepton sample); the small statistical correlation between
the $\Ds$-hadron events, the exclusively reconstructed $\mbox{B}^0_s$ mesons
and the rest of the analyses has been neglected.
The combined lower limit on $\dms$ turned out to be:
\begin{eqnarray}
   \begin{array}{ll}
\dms > 8.5 ~ \mbox{ps}^{-1}~\mbox{at the 95\% C.L.} & \\
\mbox{\rm {with a sensitivity of}}~~ \dms = 12.0 ~ \mbox{ps}^{-1}.
   \end{array}
 \end{eqnarray}
The sensitivity without systematics is of 12.5 $~\mbox{ps}^{-1}$. 
The second excluded region between $11.8$~ps$^{-1}$ and $14.0$~ps$^{-1}$  
(Figure~\ref{fig:figcomb})
is above the combined sensitivity and is not considered further.

\begin{figure}[ph] 
\begin{center}
\epsfig{figure=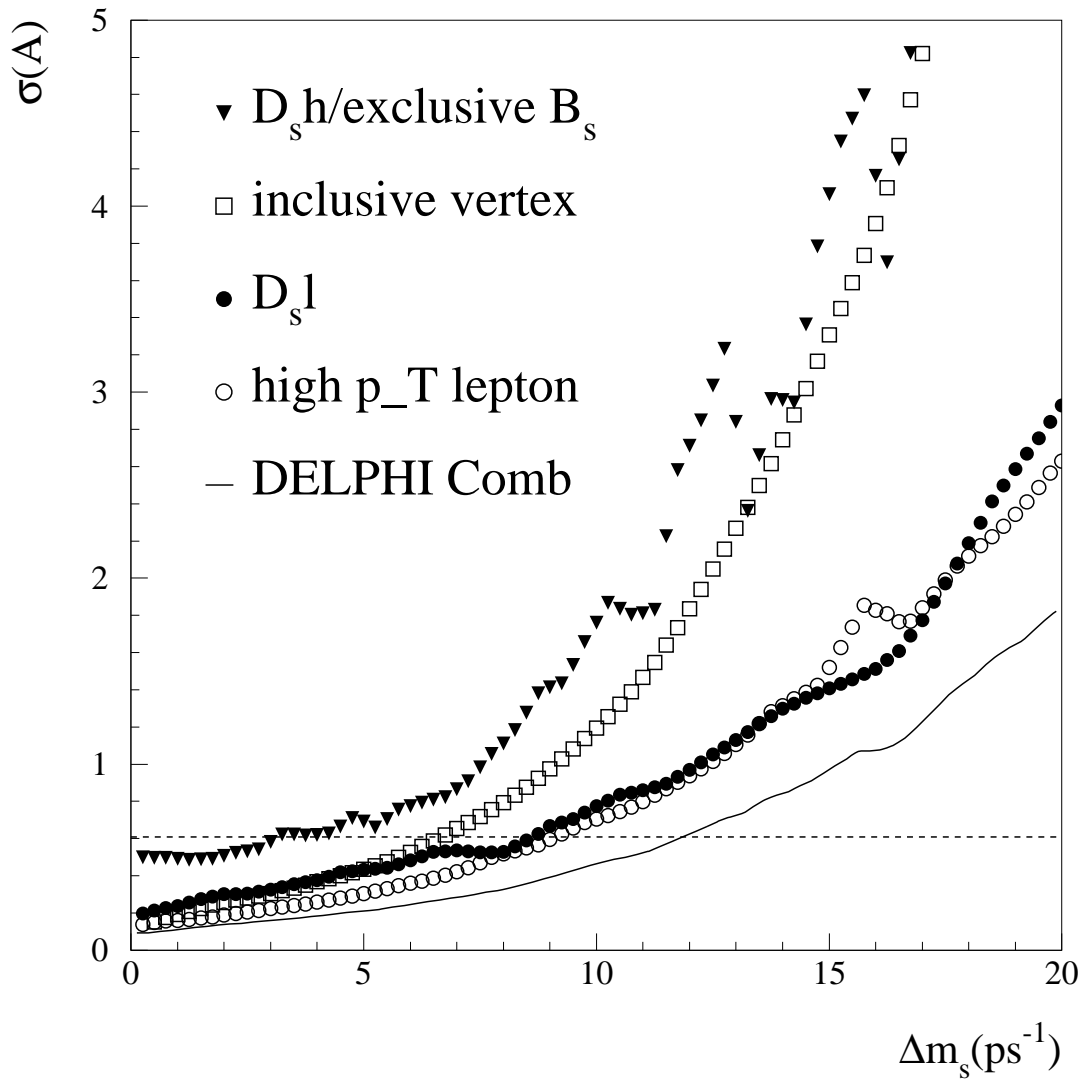,height=15cm}
\caption{ \it { Measured uncertainty on the amplitude as a function of $\Delta m_{s}$ for 
all DELPHI analyses. For reference the line $\sigma=1/1.645$ is also drawn: the 
   abscissa of the intersection with each error curve is the sensitivity of the
   analysis.}}
\label{fig:errcomb}
\end{center}
\end{figure}

The variation of the uncertainty on the amplitude as a function of $\dms$ 
for all the analyses is given in Figure~\ref{fig:errcomb}.

\pagebreak

\subsection*{Acknowledgements}
\vskip 3 mm
 We are greatly indebted to our technical 
collaborators, to the members of the CERN-SL Division for the excellent 
performance of the LEP collider, and to the funding agencies for their
support in building and operating the DELPHI detector.\\
We acknowledge in particular the support of \\
Austrian Federal Ministry of Education, Science and Culture,
GZ 616.364/2-III/2a/98, \\
FNRS--FWO, Flanders Institute to encourage scientific and technological 
research in the industry (IWT), Federal Office for Scientific, Technical
and Cultural affairs (OSTC), Belgium,  \\
FINEP, CNPq, CAPES, FUJB and FAPERJ, Brazil, \\
Czech Ministry of Industry and Trade, GA CR 202/99/1362,\\
Commission of the European Communities (DG XII), \\
Direction des Sciences de la Mati$\grave{\mbox{\rm e}}$re, CEA, France, \\
Bundesministerium f$\ddot{\mbox{\rm u}}$r Bildung, Wissenschaft, Forschung 
und Technologie, Germany,\\
General Secretariat for Research and Technology, Greece, \\
National Science Foundation (NWO) and Foundation for Research on Matter (FOM),
The Netherlands, \\
Norwegian Research Council,  \\
State Committee for Scientific Research, Poland, SPUB-M/CERN/PO3/DZ296/2000,
SPUB-M/CERN/PO3/DZ297/2000 and 2P03B 104 19 and 2P03B 69 23(2002-2004)\\
JNICT--Junta Nacional de Investiga\c{c}\~{a}o Cient\'{\i}fica 
e Tecnol$\acute{\mbox{\rm o}}$gica, Portugal, \\
Vedecka grantova agentura MS SR, Slovakia, Nr. 95/5195/134, \\
Ministry of Science and Technology of the Republic of Slovenia, \\
CICYT, Spain, AEN99-0950 and AEN99-0761,  \\
The Swedish Natural Science Research Council,      \\
Particle Physics and Astronomy Research Council, UK, \\
Department of Energy, USA, DE-FG02-01ER41155, \\
EEC RTN contract HPRN-CT-00292-2002. \\

\pagebreak

\pagebreak

\pagebreak

\pagebreak

\pagebreak

\pagebreak

\pagebreak

\pagebreak

\pagebreak

\pagebreak

\pagebreak


\pagebreak


\pagebreak

\pagebreak

\pagebreak

\end{document}